\documentclass{CSML}

\def\dOi{13(2:4)2017}
\lmcsheading%
{\dOi}
{1--27}
{}
{}
{May~\phantom{.0}1, 2016}
{May\phantom.~11, 2017}
{}

\usepackage{amsmath,amssymb,amsthm}
\usepackage[noadjust]{cite}
\usepackage{color}
\definecolor{linkcolor}{RGB}{170,0,0}
\definecolor{citecolor}{RGB}{44,160,46}
\definecolor{urlcolor}{RGB}{0,51,153}
\usepackage{breakurl}
\usepackage[breaklinks]{hyperref}
\usepackage{doi}
\usepackage{isabelle,isabellesym}
\usepackage{bold-extra}
\usepackage{subfig}
\captionsetup[subfigure]{margin=0pt, parskip=0pt,%
hangindent=0pt, indention=0pt, singlelinecheck=true}
\usepackage{pifont}
\usepackage{ulem}
\normalem
\usepackage{tikz}
\usetikzlibrary{positioning,fit,arrows,decorations.markings}
\usetikzlibrary{matrix}
\usepackage{dirtree}
\usepackage{xspace}
\usepackage{booktabs}
\usepackage{mathpartir}
\usepackage{mathtools} 
\usepackage[within=none]{newfloat}
\DeclareFloatingEnvironment[placement=tbhp]{listing}

\newcommand{\quoted}[1]{``\emph{#1}''}
\newcommand{\To}{\Rightarrow}

\DeclareSymbolFont{letters}{OML}{cmbboard}{m}{it} 

\newcommand\src{\textsf{src}}
\newcommand\f[1]{\mathsf{#1}}
\newcommand\m[1]{\ensuremath{\mathsf{#1}}}
\newcommand\Pos{{\mathcal{P}\mathsf{os}}}
\newcommand\FPos{{\Pos_\mathcal{F}}}
\newcommand\VPos{{\Pos_\mathcal{V}}}
\newcommand{\from}{\leftarrow}
\renewcommand{\conv}{\leftrightarrow}

\newcommand\NUM{164\xspace}
\newcommand\SUCC{{\text{\ding{51}}}}
\newcommand\FAIL{{\text{\ding{55}}}}

\newcommand{\seq}[2][n]{{#2_1},\ldots,{#2_{#1}}}

\newcommand\Nat{\ensuremath{\mathbb{N}}\xspace}
\newcommand\Rat{\ensuremath{\mathbb{Q}}\xspace}
\newcommand\SIG[1]{\ensuremath{\mathcal{#1}}}
\newcommand\VAR[1]{\ensuremath{\mathcal{#1}}}
\newcommand\FF{\SIG{F}}
\newcommand\VV{\VAR{V}}
\newcommand\RR{\ensuremath{\mathcal{R}}\xspace}
\newcommand\RRd{\ensuremath{\RR_\textsf{d}}\xspace}
\newcommand\RRnd{\ensuremath{\RR_\textsf{nd}}\xspace}
\newcommand\RRdnd{\ensuremath{{\RRd/\RRnd}}\xspace}
\renewcommand\SS{\ensuremath{\mathcal{S}}\xspace}
\newcommand\TT{\ensuremath{\mathcal{T}}\xspace}
\newcommand\FVTERMS{\TT(\FF,\VV)}
\newcommand\REL[2]{{\ensuremath{#1 \kern0em/\kern0em #2}}\xspace}
\newcommand\RS{\REL{\RR}{\SS}}
\newcommand{\sep}{\hspace{-0.28em}}
\newcommand{\cp}{\mathrel{\from\sep\rtimes\sep\to}}
\newcommand{\cps}[1]{\mathrel{\from\sep{#1}\sep\to}}

\newcommand\ACP{\ensuremath{\mathsf{ACP}}\xspace}
\newcommand\CSI{\ensuremath{\mathsf{CSI}}\xspace}
\newcommand\ISAFOR{\textsf{Isa\kern-0.2exF\kern-0.2exo\kern-0.2exR}\xspace}
\newcommand\ISAR{\textsf{Isar}\xspace}
\newcommand\CETA{\textsf{C\kern-0.2exe\kern-0.5exT\kern-0.5exA}\xspace}
\newcommand\COQ{\textsf{Coq}\xspace}
\newcommand\ISABELLE{\textsf{Isabelle}\xspace}
\newcommand\HOL{\textsf{HOL}\xspace}
\newcommand\ACL{\textsf{ACL2}\xspace}
\newcommand\PVS{\textsf{PVS}\xspace}
\newcommand\CPF{\textsf{CPF}\xspace}
\newcommand\SAIGAWA{\ensuremath{\mathsf{saigawa}}\xspace}

\newcommand{\smallparallel}{%
 \def\next##1##2{\raise.07ex\hbox{$##1{\shortmid}\mkern-1.5mu%
{\shortmid}$}}%
 \mathpalette\next{}%
}

\newcommand{\overlayrel}[4]{%
 \mathrel{%
 \def\next##1##2{%
 \setbox0=\hbox{$##1#3$}%
 \dimen0=\wd0%
 \hbox to0pt{\box0}%
 \hbox to\dimen0{$##1\hfil\mkern#1mu#4\mkern#2mu\hfil$}}%
 \mathpalette\next{}}%
}
\newcommand{\prightarrow}{\overlayrel02\rightarrow\smallparallel}
\newcommand{\xprightarrow}[2][]{\overlayrel02{\xrightarrow[#1]{#2}}%
\smallparallel}
\makeatletter
\newcommand{\r@rrow}[3]{%
 \newcommand{#1}[2][]{%
  \def\next{#2\@ifempty{##1}{}{_{##1}}\@ifempty{##2}{}{^{##2}}}%
  \mathchoice{#3[##1]{##2}}{\next}{\next}{\next}%
 }%
}
\newcommand{\l@rrow}[3]{%
 \newcommand{#1}[2][]{%
  \def\next####1{%
   \setbox0=\hbox{$####1\vphantom{#2}\@ifempty{##1}{}{_{}}%
    \@ifempty{##2}{}{^{##2}}$}%
   \setbox1=\hbox{$####1\vphantom{#2}\@ifempty{##1}{}{_{##1}}%
    \@ifempty{##2}{}{^{}}$}%
   \setbox2=\vbox{\hbox to\wd0{}\hbox to\wd1{}}%
   \mathrel{\hskip\wd2\hskip-\wd0\box0\hskip-\wd1\box1{#2}}%
  }%
  \mathchoice{#3[##1]{##2}}{\next\textstyle}{\next\scriptstyle}%
 {\next\scriptscriptstyle}%
 }%
}

\r@rrow{\xc}{\leftrightarrow}{\xrightarrow}
\r@rrow{\xr}{\rightarrow}{\xrightarrow}
\l@rrow{\xl}{\leftarrow}{\xleftarrow}
\r@rrow{\xlr}{\leftrightarrow}{\xleftrightarrow}
\r@rrow{\xR}{\prightarrow}{\xprightarrow}
\r@rrow{\Xr}{\Rightarrow}{\xRightarrow}
\l@rrow{\Xl}{\Leftarrow}{\xLeftarrow}
\r@rrow{\Xlr}{\Leftrightarrow}{\xLeftrightarrow}

\newcommand{\turnleft}[1]{%
 \def\next##1##2{\raisebox{-.09em}{\rotatebox{90}{$##1#1$}}}%
 \mathpalette\next{}%
}
\newcommand{\rel}{\succcurlyeq}
\newcommand{\reli}{\preccurlyeq}
\newcommand{\lablt}{<}
\newcommand{\lable}{\leqslant}
\newcommand{\Vee}{\turnleft\lablt}
\newcommand{\Veq}{\turnleft\lable}
\newcommand{\Vrel}{\turnleft{\reli}}

\tikzset{
to/.style={->},
bi/.style={<->},
To/.style={->,postaction={decorate},
 decoration={markings,
 mark=at position .5 with
 \node[transform shape]{$\large\mathstrut\raisebox{0.02em}%
 {$\smallparallel$}$};
}},
TO/.style={->,postaction={decorate},
 decoration={markings,
 mark=at position .5 with
 \node[transform shape]{$\mathstrut\smalldevel$};
}},
tos/.style={->>},
hide/.style={color=white},
imp/.style={-implies,double},
}

\newcommand{\vl}[1]{\acute{#1}}
\newcommand{\vlr}[1]{\hat{#1}}
\newcommand{\vr}[1]{\grave{#1}}
\newcommand{\vu}[1]{\bar{#1}}

\isabellestyleliteral
\renewcommand\isastyle{\isastyleminor}

\newcommand{\isaname}[1]{\textcolor{blue!70!black}{\textbf{#1\/}}}
\newcommand{\isaconst}[1]{\textcolor{blue!70!black}{\emph{\textsf{#1}}}}
\renewcommand{\isakeyword}[1]{\textbf{#1}}
\newcommand{\isasubterm}{$|\underline{\ }$}

\newcommand{\DefineSnippet}[2]{%
  \expandafter\newcommand\csname snippet--#1\endcsname{#2}}
\newcommand\Snippet[1]{%
  \begin{isabelle}
  \csname snippet--#1\endcsname
  \end{isabelle}}
\newcommand\snippet[1]{{%
  {\isastyle\csname snippet--#1\endcsname}%
}}

\DefineSnippet{peaks}{%
\begin{isabelle}%
\isaname{local{\isacharunderscore}peaks}\ {\isasymR}\ {\isacharequal}\isanewline
\isaindent{\ \ \ \ }{\isacharbraceleft}{\isacharparenleft}{\isacharparenleft}s{\isacharcomma}\ r\isactrlsub {\isadigit{1}}{\isacharcomma}\ p\isactrlsub {\isadigit{1}}{\isacharcomma}\ {\isasymsigma}\isactrlsub {\isadigit{1}}{\isacharcomma}\ \isaconst{True}{\isacharcomma}\ t{\isacharparenright}{\isacharcomma}\ {\isacharparenleft}s{\isacharcomma}\ r\isactrlsub {\isadigit{2}}{\isacharcomma}\ p\isactrlsub {\isadigit{2}}{\isacharcomma}\ {\isasymsigma}\isactrlsub {\isadigit{2}}{\isacharcomma}\ \isaconst{True}{\isacharcomma}\ u{\isacharparenright}{\isacharparenright}\ {\isacharbar}s\ t\ u\ r\isactrlsub {\isadigit{1}}\ r\isactrlsub {\isadigit{2}}\ p\isactrlsub {\isadigit{1}}\ p\isactrlsub {\isadigit{2}}\ {\isasymsigma}\isactrlsub {\isadigit{1}}\ {\isasymsigma}\isactrlsub {\isadigit{2}}{\isachardot}\isanewline
\isaindent{\ \ \ \ \ }{\isacharparenleft}s{\isacharcomma}\ t{\isacharparenright}\ {\isasymin}\ \isaname{rstep{\isacharunderscore}r{\isacharunderscore}p{\isacharunderscore}s}\ {\isasymR}\ r\isactrlsub {\isadigit{1}}\ p\isactrlsub {\isadigit{1}}\ {\isasymsigma}\isactrlsub {\isadigit{1}}\ \isasymand\ {\isacharparenleft}s{\isacharcomma}\ u{\isacharparenright}\ {\isasymin}\ \isaname{rstep{\isacharunderscore}r{\isacharunderscore}p{\isacharunderscore}s}\ {\isasymR}\ r\isactrlsub {\isadigit{2}}\ p\isactrlsub {\isadigit{2}}\ {\isasymsigma}\isactrlsub {\isadigit{2}}{\isacharbraceright}%
\end{isabelle}
\begin{isabelle}%
\isaname{parallel{\isacharunderscore}peak}\ {\isasymR}\ p\ {\isacharequal}\isanewline
\isaindent{\ \ \ \ }{\isacharparenleft}p\ {\isasymin}\ \isaname{local{\isacharunderscore}peaks}\ {\isasymR}\ \isasymand\isanewline
\isaindent{\ \ \ \ \ }{\isacharparenleft}\isakeyword{let}\ {\isacharparenleft}{\isacharparenleft}s{\isacharcomma}\ r\isactrlsub {\isadigit{1}}{\isacharcomma}\ p\isactrlsub {\isadigit{1}}{\isacharcomma}\ {\isasymsigma}\isactrlsub {\isadigit{1}}{\isacharcomma}\ b{\isacharcomma}\ t{\isacharparenright}{\isacharcomma}\ {\isacharparenleft}s{\isacharcomma}\ r\isactrlsub {\isadigit{2}}{\isacharcomma}\ p\isactrlsub {\isadigit{2}}{\isacharcomma}\ {\isasymsigma}\isactrlsub {\isadigit{2}}{\isacharcomma}\ d{\isacharcomma}\ u{\isacharparenright}{\isacharparenright}\ {\isacharequal}\ p\ \isakeyword{in}\ p\isactrlsub {\isadigit{1}}\ {\isasymbottom}\ p\isactrlsub {\isadigit{2}}{\isacharparenright}{\isacharparenright}%
\end{isabelle}
\begin{isabelle}%
\isaname{function{\isacharunderscore}peak}\ {\isasymR}\ p\ {\isacharequal}\isanewline
\isaindent{\ \ \ \ }{\isacharparenleft}p\ {\isasymin}\ \isaname{local{\isacharunderscore}peaks}\ {\isasymR}\ \isasymand\isanewline
\isaindent{\ \ \ \ \ }{\isacharparenleft}\isakeyword{let}\ {\isacharparenleft}{\isacharparenleft}s{\isacharcomma}\ rl\isactrlsub {\isadigit{1}}{\isacharcomma}\ p\isactrlsub {\isadigit{1}}{\isacharcomma}\ {\isasymsigma}\isactrlsub {\isadigit{1}}{\isacharcomma}\ b{\isacharcomma}\ t{\isacharparenright}{\isacharcomma}\ {\isacharparenleft}s{\isacharcomma}\ rl\isactrlsub {\isadigit{2}}{\isacharcomma}\ p\isactrlsub {\isadigit{2}}{\isacharcomma}\ {\isasymsigma}\isactrlsub {\isadigit{2}}{\isacharcomma}\ d{\isacharcomma}\ u{\isacharparenright}{\isacharparenright}\ {\isacharequal}\ p\ \isakeyword{in}\isanewline
\isaindent{\ \ \ \ \ {\isacharparenleft}}{\isasymexists}r{\isachardot}\ p\isactrlsub {\isadigit{1}}\ {\isacharless}{\isacharhash}{\isachargreater}\ r\ {\isacharequal}\ p\isactrlsub {\isadigit{2}}\ \isasymand\ r\ {\isasymin}\ \isaname{poss}\ {\isacharparenleft}\isaname{fst}\ rl\isactrlsub {\isadigit{1}}{\isacharparenright}\ \isasymand\ \isaname{is{\isacharunderscore}Fun}\ {\isacharparenleft}\isaname{fst}\ rl\isactrlsub {\isadigit{1}}\ \isasubterm\ r{\isacharparenright}{\isacharparenright}{\isacharparenright}%
\end{isabelle}
\begin{isabelle}%
\isaname{variable{\isacharunderscore}peak}\ {\isasymR}\ p\ {\isacharequal}\isanewline
\isaindent{\ \ \ \ }{\isacharparenleft}p\ {\isasymin}\ \isaname{local{\isacharunderscore}peaks}\ {\isasymR}\ \isasymand\isanewline
\isaindent{\ \ \ \ \ }{\isacharparenleft}\isakeyword{let}\ {\isacharparenleft}{\isacharparenleft}s{\isacharcomma}\ rl\isactrlsub {\isadigit{1}}{\isacharcomma}\ p\isactrlsub {\isadigit{1}}{\isacharcomma}\ {\isasymsigma}\isactrlsub {\isadigit{1}}{\isacharcomma}\ b{\isacharcomma}\ t{\isacharparenright}{\isacharcomma}\ {\isacharparenleft}s{\isacharcomma}\ rl\isactrlsub {\isadigit{2}}{\isacharcomma}\ p\isactrlsub {\isadigit{2}}{\isacharcomma}\ {\isasymsigma}\isactrlsub {\isadigit{2}}{\isacharcomma}\ d{\isacharcomma}\ u{\isacharparenright}{\isacharparenright}\ {\isacharequal}\ p\ \isakeyword{in}\isanewline
\isaindent{\ \ \ \ \ {\isacharparenleft}}{\isasymexists}r{\isachardot}\ p\isactrlsub {\isadigit{1}}\ {\isacharless}{\isacharhash}{\isachargreater}\ r\ {\isacharequal}\ p\isactrlsub {\isadigit{2}}\ \isasymand\ {\isasymnot}\ {\isacharparenleft}r\ {\isasymin}\ \isaname{poss}\ {\isacharparenleft}\isaname{fst}\ rl\isactrlsub {\isadigit{1}}{\isacharparenright}\ \isasymand\ \isaname{is{\isacharunderscore}Fun}\ {\isacharparenleft}\isaname{fst}\ rl\isactrlsub {\isadigit{1}}\ \isasubterm\ r{\isacharparenright}{\isacharparenright}{\isacharparenright}{\isacharparenright}%
\end{isabelle}
}
\DefineSnippet{peaks_cases}{%
\begin{isabelle}%
p\ {\isasymin}\ \isaname{local{\isacharunderscore}peaks}\ {\isasymR}\ {\isasymLongrightarrow}\isanewline
\isaname{parallel{\isacharunderscore}peak}\ {\isasymR}\ p\ \isasymor\ \isaname{variable{\isacharunderscore}peak}\ {\isasymR}\ p\ \isasymor\ \isaname{variable{\isacharunderscore}peak}\ {\isasymR}\ {\isacharparenleft}\isaname{snd}\ p{\isacharcomma}\ \isaname{fst}\ p{\isacharparenright}\ \isasymor\isanewline
\isaname{function{\isacharunderscore}peak}\ {\isasymR}\ p\ \isasymor\ \isaname{function{\isacharunderscore}peak}\ {\isasymR}\ {\isacharparenleft}\isaname{snd}\ p{\isacharcomma}\ \isaname{fst}\ p{\isacharparenright}%
\end{isabelle}
}
\DefineSnippet{ldc_trs_def}{%
\begin{isabelle}%
\isaname{ldc{\isacharunderscore}trs}\ R\ p\ cl\isactrlsub {\isadigit{1}}\ sl\ cl\isactrlsub {\isadigit{2}}\ cr\isactrlsub {\isadigit{1}}\ sr\ cr\isactrlsub {\isadigit{2}}\ {\isacharequal}\isanewline
\isaindent{\ \ \ \ }{\isacharparenleft}p\ {\isasymin}\ \isaname{local{\isacharunderscore}peaks}\ R\ \isasymand\ cl\isactrlsub {\isadigit{1}}\ {\isasymin}\ \isaname{conv}\ R\ \isasymand\ sl\ {\isasymin}\ \isaname{seq}\ R\ \isasymand\ cl\isactrlsub {\isadigit{2}}\ {\isasymin}\ \isaname{conv}\ R\ \isasymand\isanewline
\isaindent{\ \ \ \ \ }cr\isactrlsub {\isadigit{1}}\ {\isasymin}\ \isaname{conv}\ R\ \isasymand\ sr\ {\isasymin}\ \isaname{seq}\ R\ \isasymand\ cr\isactrlsub {\isadigit{2}}\ {\isasymin}\ \isaname{conv}\ R\ \isasymand\isanewline
\isaindent{\ \ \ \ \ }\isaname{get{\isacharunderscore}target}\ {\isacharparenleft}\isaname{fst}\ p{\isacharparenright}\ {\isacharequal}\ \isaname{first}\ cl\isactrlsub {\isadigit{1}}\ \isasymand\ \isaname{last}\ cl\isactrlsub {\isadigit{1}}\ {\isacharequal}\ \isaname{first}\ sl\ \isasymand\isanewline
\isaindent{\ \ \ \ \ }\isaname{last}\ sl\ {\isacharequal}\ \isaname{first}\ cl\isactrlsub {\isadigit{2}}\ \isasymand\ \isaname{get{\isacharunderscore}target}\ {\isacharparenleft}\isaname{snd}\ p{\isacharparenright}\ {\isacharequal}\ \isaname{first}\ cr\isactrlsub {\isadigit{1}}\ \isasymand\isanewline
\isaindent{\ \ \ \ \ }\isaname{last}\ cr\isactrlsub {\isadigit{1}}\ {\isacharequal}\ \isaname{first}\ sr\ \isasymand\ \isaname{last}\ sr\ {\isacharequal}\ \isaname{first}\ cr\isactrlsub {\isadigit{2}}\ \isasymand\ \isaname{last}\ cl\isactrlsub {\isadigit{2}}\ {\isacharequal}\ \isaname{last}\ cr\isactrlsub {\isadigit{2}}{\isacharparenright}%
\end{isabelle}
}
\DefineSnippet{eldc_def}{%
\begin{isabelle}%
\isaname{eldc}\ {\isasymR}\ $\ell$\ r\ p\ {\isacharequal}\isanewline
\isaindent{\ \ \ \ }{\isacharparenleft}{\isasymexists}cl\isactrlsub {\isadigit{1}}\ sl\ cl\isactrlsub {\isadigit{2}}\ cr\isactrlsub {\isadigit{1}}\ sr\ cr\isactrlsub {\isadigit{2}}{\isachardot}\ \isaname{ldc{\isacharunderscore}trs}\ {\isasymR}\ p\ cl\isactrlsub {\isadigit{1}}\ sl\ cl\isactrlsub {\isadigit{2}}\ cr\isactrlsub {\isadigit{1}}\ sr\ cr\isactrlsub {\isadigit{2}}\ \isasymand\isanewline
\isaindent{\ \ \ \ {\isacharparenleft}\ \ \ }\isaname{eld{\isacharunderscore}conv}\ $\ell$\ r\ p\ cl\isactrlsub {\isadigit{1}}\ sl\ cl\isactrlsub {\isadigit{2}}\ cr\isactrlsub {\isadigit{1}}\ sr\ cr\isactrlsub {\isadigit{2}}{\isacharparenright}%
\end{isabelle}
}
\DefineSnippet{eld_conv_def}{%
\begin{isabelle}%
\isaname{eld{\isacharunderscore}conv}\ $\ell$\ r\ p\ cl\isactrlsub {\isadigit{1}}\ sl\ cl\isactrlsub {\isadigit{2}}\ cr\isactrlsub {\isadigit{1}}\ sr\ cr\isactrlsub {\isadigit{2}}\ {\isacharequal}\isanewline
\isaindent{\ \ \ \ }{\isacharparenleft}\isaname{ELD{\isacharunderscore}{\isadigit{1}}}\ r\ {\isacharparenleft}$\ell$\ {\isacharparenleft}\isaname{fst}\ p{\isacharparenright}{\isacharparenright}\ {\isacharparenleft}$\ell$\ {\isacharparenleft}\isaname{snd}\ p{\isacharparenright}{\isacharparenright}\ {\isacharparenleft}\isaname{map}\ $\ell$\ {\isacharparenleft}\isaname{snd}\ cl\isactrlsub {\isadigit{1}}{\isacharparenright}{\isacharparenright}\ {\isacharparenleft}\isaname{map}\ $\ell$\ {\isacharparenleft}\isaname{snd}\ sl{\isacharparenright}{\isacharparenright}\isanewline
\isaindent{\ \ \ \ {\isacharparenleft}\ }{\isacharparenleft}\isaname{map}\ $\ell$\ {\isacharparenleft}\isaname{snd}\ cl\isactrlsub {\isadigit{2}}{\isacharparenright}{\isacharparenright}\ \isasymand\isanewline
\isaindent{\ \ \ \ \ }\isaname{ELD{\isacharunderscore}{\isadigit{1}}}\ r\ {\isacharparenleft}$\ell$\ {\isacharparenleft}\isaname{snd}\ p{\isacharparenright}{\isacharparenright}\ {\isacharparenleft}$\ell$\ {\isacharparenleft}\isaname{fst}\ p{\isacharparenright}{\isacharparenright}\ {\isacharparenleft}\isaname{map}\ $\ell$\ {\isacharparenleft}\isaname{snd}\ cr\isactrlsub {\isadigit{1}}{\isacharparenright}{\isacharparenright}\ {\isacharparenleft}\isaname{map}\ $\ell$\ {\isacharparenleft}\isaname{snd}\ sr{\isacharparenright}{\isacharparenright}\isanewline
\isaindent{\ \ \ \ \ \ }{\isacharparenleft}\isaname{map}\ $\ell$\ {\isacharparenleft}\isaname{snd}\ cr\isactrlsub {\isadigit{2}}{\isacharparenright}{\isacharparenright}{\isacharparenright}%
\end{isabelle}
}
\DefineSnippet{conv}{%
  \[\isa{\mbox{}\inferrule{\mbox{}}{\mbox{{\isacharparenleft}s{\isacharcomma}\ {\isacharbrackleft}{\isacharbrackright}{\isacharparenright}\ {\isasymin}\ \isaname{conv}\ {\isasymR}}}}\]

  \[\isa{\mbox{}\inferrule{\mbox{{\isacharparenleft}s{\isacharcomma}\ t{\isacharparenright}\ {\isasymin}\ \isaname{rstep{\isacharunderscore}r{\isacharunderscore}p{\isacharunderscore}s}\ {\isasymR}\ r\ p\ {\isasymsigma}}\\\ \mbox{{\isacharparenleft}t{\isacharcomma}\ ts{\isacharparenright}\ {\isasymin}\ \isaname{conv}\ {\isasymR}}}{\mbox{{\isacharparenleft}s{\isacharcomma}\ {\isacharparenleft}s{\isacharcomma}\ r{\isacharcomma}\ p{\isacharcomma}\ {\isasymsigma}{\isacharcomma}\ \isaconst{True}{\isacharcomma}\ t{\isacharparenright}\ {\isacharhash}\ ts{\isacharparenright}\ {\isasymin}\ \isaname{conv}\ {\isasymR}}}}\]

  \[\isa{\mbox{}\inferrule{\mbox{{\isacharparenleft}s{\isacharcomma}\ t{\isacharparenright}\ {\isasymin}\ \isaname{rstep{\isacharunderscore}r{\isacharunderscore}p{\isacharunderscore}s}\ {\isasymR}\ r\ p\ {\isasymsigma}}\\\ \mbox{{\isacharparenleft}s{\isacharcomma}\ ts{\isacharparenright}\ {\isasymin}\ \isaname{conv}\ {\isasymR}}}{\mbox{{\isacharparenleft}t{\isacharcomma}\ {\isacharparenleft}s{\isacharcomma}\ r{\isacharcomma}\ p{\isacharcomma}\ {\isasymsigma}{\isacharcomma}\ \isaconst{False}{\isacharcomma}\ t{\isacharparenright}\ {\isacharhash}\ ts{\isacharparenright}\ {\isasymin}\ \isaname{conv}\ {\isasymR}}}}\]
}
\DefineSnippet{s_to_t}{%
  \isa{{\isacharparenleft}s{\isacharcomma}\ t{\isacharparenright}\ {\isasymin}\ \isaname{rstep{\isacharunderscore}r{\isacharunderscore}p{\isacharunderscore}s}\ {\isasymR}\ {\isacharparenleft}l{\isacharcomma}\ r{\isacharparenright}\ p\ {\isasymsigma}}%
}
\DefineSnippet{rstep}{%
  \isa{\isaname{rstep{\isacharunderscore}r{\isacharunderscore}p{\isacharunderscore}s}}%
}
\DefineSnippet{eldc}{%
  \isa{\isaname{eldc}}%
}
\DefineSnippet{ldc_trs}{%
  \isa{\isaname{ldc{\isacharunderscore}trs}}%
}
\DefineSnippet{eld_conv}{%
  \isa{\isaname{eld{\isacharunderscore}conv}}%
}
\DefineSnippet{ELD_1_def}{%
\begin{isabelle}%
\isaname{ELD{\isacharunderscore}{\isadigit{1}}}\ r\ {\isasymbeta}\ {\isasymalpha}\ {\isasymsigma}\isactrlsub {\isadigit{1}}\ {\isasymsigma}\isactrlsub {\isadigit{2}}\ {\isasymsigma}\isactrlsub {\isadigit{3}}\ {\isacharequal}\isanewline
\isaindent{\ \ \ \ }{\isacharparenleft}\isaname{set}\ {\isasymsigma}\isactrlsub {\isadigit{1}}\ {\isasymsubseteq}\ \isaname{ds}\ {\isacharparenleft}\isaname{fst}\ r{\isacharparenright}\ {\isacharbraceleft}{\isasymbeta}{\isacharbraceright}\ \isasymand\ \isaname{length}\ {\isasymsigma}\isactrlsub {\isadigit{2}}\ {\isasymle}\ {\isadigit{1}}\ \isasymand\ \isaname{set}\ {\isasymsigma}\isactrlsub {\isadigit{2}}\ {\isasymsubseteq}\ \isaname{ds}\ {\isacharparenleft}\isaname{snd}\ r{\isacharparenright}\ {\isacharbraceleft}{\isasymalpha}{\isacharbraceright}\ \isasymand\isanewline
\isaindent{\ \ \ \ \ }\isaname{set}\ {\isasymsigma}\isactrlsub {\isadigit{3}}\ {\isasymsubseteq}\ \isaname{ds}\ {\isacharparenleft}\isaname{fst}\ r{\isacharparenright}\ {\isacharbraceleft}{\isasymalpha}{\isacharcomma}\ {\isasymbeta}{\isacharbraceright}{\isacharparenright}%
\end{isabelle}%
}
\DefineSnippet{ds}{%
  \isa{\isaname{ds}}%
}
\DefineSnippet{ELD_1}{%
  \isa{\isaname{ELD{\isacharunderscore}{\isadigit{1}}}}%
}
\DefineSnippet{cl1}{%
  \isa{cl\isactrlsub {\isadigit{1}}}%
}
\DefineSnippet{cl2}{%
  \isa{cl\isactrlsub {\isadigit{2}}}%
}
\DefineSnippet{cr1}{%
  \isa{cr\isactrlsub {\isadigit{1}}}%
}
\DefineSnippet{cr2}{%
  \isa{cr\isactrlsub {\isadigit{2}}}%
}
\DefineSnippet{sl}{%
  \isa{sl}%
}
\DefineSnippet{sr}{%
  \isa{sr}%
}
\DefineSnippet{True}{%
  \isa{\isaconst{True}}%
}
\DefineSnippet{False}{%
  \isa{\isaconst{False}}%
}
\DefineSnippet{local_peaks}{%
  \isa{\isaname{local{\isacharunderscore}peaks}}%
}
\DefineSnippet{function_peak}{%
  \isa{\isaname{function{\isacharunderscore}peak}}%
}
\DefineSnippet{variable_peak}{%
  \isa{\isaname{variable{\isacharunderscore}peak}}%
}
\DefineSnippet{first}{%
  \isa{\isaname{first}}%
}
\DefineSnippet{last}{%
  \isa{\isaname{last}}%
}
\DefineSnippet{get_target}{%
  \isa{\isaname{get{\isacharunderscore}target}}%
}
\DefineSnippet{greek_less_def}{%
\begin{isabelle}%
\isaname{adj{\isacharunderscore}msog}\ xs\ zs\ l\ {\isacharequal}\isanewline
\isaindent{\ \ \ \ }{\isacharparenleft}\isakeyword{case}\ l\ \isakeyword{of}\ {\isacharparenleft}y{\isacharcomma}\ ys{\isacharparenright}\ {\isasymRightarrow}\isanewline
\isaindent{\ \ \ \ {\isacharparenleft}\isakeyword{case}\ l\ \isakeyword{of}\ \ \ }{\isacharparenleft}y{\isacharcomma}\ \isakeyword{case}\ \isaname{fst}\ y\ \isakeyword{of}\ \isaconst{Acute}\ {\isasymRightarrow}\ ys\ \isacharat\ zs\ {\isacharbar}\ \isaconst{Grave}\ {\isasymRightarrow}\ xs\ \isacharat\ ys\ {\isacharbar}\ \isaconst{Macron}\ {\isasymRightarrow}\ ys{\isacharparenright}{\isacharparenright}%
\end{isabelle}%
\begin{isabelle}%
\isaname{ms{\isacharunderscore}of{\isacharunderscore}greek}\ as\ {\isacharequal}\isanewline
\isaindent{\ \ \ \ }\isaname{mset}\ {\isacharparenleft}\isaname{map}\ {\isacharparenleft}{\isasymlambda}x{\isachardot}\ \isakeyword{case}\ x\ \isakeyword{of}\ {\isacharparenleft}xs{\isacharcomma}\ y{\isacharcomma}\ zs{\isacharparenright}\ {\isasymRightarrow}\ \isaname{adj{\isacharunderscore}msog}\ xs\ zs\ {\isacharparenleft}y{\isacharcomma}\ {\isacharbrackleft}{\isacharbrackright}{\isacharparenright}{\isacharparenright}\ {\isacharparenleft}\isaname{list{\isacharunderscore}splits}\ as{\isacharparenright}{\isacharparenright}%
\end{isabelle}%
\begin{isabelle}%
\isaname{letter{\isacharunderscore}less}\ r\ {\isacharequal}\ {\isacharbraceleft}{\isacharparenleft}a{\isacharcomma}\ b{\isacharparenright}{\isachardot}\ {\isacharparenleft}\isaname{snd}\ a{\isacharcomma}\ \isaname{snd}\ b{\isacharparenright}\ {\isasymin}\ r{\isacharbraceright}%
\end{isabelle}%
\begin{isabelle}%
\isaname{nest}\ r\ s\ {\isacharequal}\ {\isacharbraceleft}{\isacharparenleft}a{\isacharcomma}\ b{\isacharparenright}{\isachardot}\ {\isacharparenleft}\isaname{ms{\isacharunderscore}of{\isacharunderscore}greek}\ a{\isacharcomma}\ \isaname{ms{\isacharunderscore}of{\isacharunderscore}greek}\ b{\isacharparenright}\ {\isasymin}\ \isaname{mult}\ {\isacharparenleft}\isaname{letter{\isacharunderscore}less}\ r\ {\isacharless}{\isacharasterisk}lex{\isacharasterisk}{\isachargreater}\ s{\isacharparenright}{\isacharbraceright}%
\end{isabelle}%
\begin{isabelle}%
\isaname{greek{\isacharunderscore}less}\ r\ {\isacharequal}\ \isaname{lfp}\ {\isacharparenleft}\isaname{nest}\ r{\isacharparenright}%
\end{isabelle}%
}
\DefineSnippet{greek_less}{%
  \isa{\isaname{greek{\isacharunderscore}less}}%
}
\DefineSnippet{adj_msog}{%
  \isa{\isaname{adj{\isacharunderscore}msog}}%
}
\DefineSnippet{ms_of_greek}{%
  \isa{\isaname{ms{\isacharunderscore}of{\isacharunderscore}greek}}%
}
\DefineSnippet{letter_less}{%
  \isa{\isaname{letter{\isacharunderscore}less}}%
}
\DefineSnippet{nest}{%
  \isa{\isaname{nest}}%
}
\DefineSnippet{mult}{%
  \isa{\isaname{mult}}%
}
\DefineSnippet{lfp}{%
  \isa{\isaname{lfp}}%
}
\DefineSnippet{list_splits}{%
  \isa{\isaname{list{\isacharunderscore}splits}}%
}

\begin{document}

\title[Certifying Confluence Proofs via Relative Termination and Rule Labeling]{Certifying Confluence Proofs via Relative Termination and Rule Labeling\rsuper*}

\author[J.~Nagele]{Julian Nagele\rsuper a}
\address{{\lsuper{a,b}}Department of Computer Science\\University of Innsbruck\\Austria}
\email{\{julian.nagele, bertram.felgenhauer\}@uibk.ac.at}
\thanks{{\lsuper{a,b}}This research is supported by FWF (Austrian Science Fund) project
P27528.}

\author[B.~Felgenhauer]{Bertram Felgenhauer\rsuper b}
\address{\vspace{-18 pt}}

\author[H.~Zankl]{Harald Zankl\rsuper c}
\address{{\lsuper c}Innsbruck\\Austria}
\email{hzankl@gmail.com}
\titlecomment{{\lsuper*}A preliminary version of this article appeared in RTA 2015.}


\begin{abstract}
\noindent
The rule labeling heuristic aims to establish confluence of (left-)linear
term rewrite systems via decreasing diagrams. We present a formalization of a
confluence criterion based on the interplay of relative termination and 
the rule labeling in the theorem prover Isabelle. Moreover, we report on 
the integration of this result into the certifier CeTA, facilitating the 
checking of confluence certificates based on decreasing diagrams.
The power of the method is illustrated by an experimental
evaluation on a (standard) collection of confluence problems.
\end{abstract}

\maketitle

\section{Introduction}

One of the most important properties in program verification is confluence,
which ensures that different computation paths produce the same result, i.e.,
that normal forms are unique.  Consequently automatable formal methods that
can analyze confluence of programs are of great interest.  One of the most
widely used and successful formalisms for analyzing confluence is 
rewriting, a conceptually simple but powerful abstract model of computation,
which uses directed equations, i.e., rewrite rules to replace equals by equals.
The recent advances in confluence research have culminated in
the Confluence Competition (CoCo)~\cite{AHNNZ15},
where automated tools try to establish/refute confluence of rewrite systems.%
\footnote{\label{FOO:coco}\url{http://coco.nue.riec.tohoku.ac.jp}}
As demonstrated in the termination competition,%
\footnote{\url{http://termination-portal.org}}
rewriting is suitable for analyzing programs written in real-world programming
languages such as C, Haskell, and Java.
As the proofs produced by automated tools are often complex and large, there is
interest in verifying them using an independent certifier.
A certifier is a different kind of automated tool that reads proof
certificates, which are for instance produced by
automated confluence tools, and either
accepts them as correct or rejects them as erroneous, thereby increasing
credibility of the proofs found by the confluence tools.
To ensure correctness of
the certifier itself, the predominant solution is to use proof assistants like
\COQ or \ISABELLE to first formalize the underlying theory in the proof
assistant and then use this formalization to obtain verified, executable check
functions for inspecting the certificates.

\begin{figure}[t]
\centering
\begin{tikzpicture}[semithick]
  \node (Lit) at (0,0) [minimum height=7mm,draw,minimum width=3cm] {Literature};
  \node (CT) at (9.5,0) [minimum height=7mm,draw,minimum width=3cm] {Confluence Tool};
  \path (Lit) edge[-stealth] node[above] {implementation} (CT);
  \path (9.5,1) edge[dashed,-stealth] node[left] {rewrite sys.} (CT);
  \node (Proof) at (9.5,-2) {Proof};
  \node [left of=Proof, node distance=1cm] {\CPF};
  \path (CT) edge[dashed] (Proof);
  \node (Isabelle) at (0,-2) [rounded corners=3.5mm,minimum height=7mm, draw,minimum width=3.0cm] {\ISABELLE/\HOL};
  \node (Isafor) at (0,-4) [minimum height=7mm,draw,minimum width=3cm] {\ISAFOR};
  \node (Ceta) at (9.5,-4) [minimum height=7mm,draw,minimum width=3cm] {\CETA};
  \path (Lit) edge node[right,text width=2cm] {theorems \& proofs} (Isabelle);
  \path (Isabelle) edge[-stealth] (Isafor);
  \path (Isafor) edge[-stealth] node[above] {code generation \&} node[below] {Haskell compiler} (Ceta);
  \path (Ceta) edge[dashed,-stealth] node[left] {accept/reject} (9.5,-5);
  \path (Proof) edge[dashed,-stealth] (Ceta);
  \node (form) at (0,-2) [rounded corners=3mm, draw, minimum height=6.25cm, minimum width=4.9cm, dashed] {};
  \node (formtext) at (0,-5.5) {Formalization};
  \node (cert) at (9.5,-2) [rounded corners=3mm, draw, minimum height=6.25cm, minimum width=4.9cm, dashed] {};
  \node (certtext) at (9.5,-5.5) {Certification};
\end{tikzpicture}
\caption{Certification of confluence proofs.}
\label{fig:certification}
\end{figure}
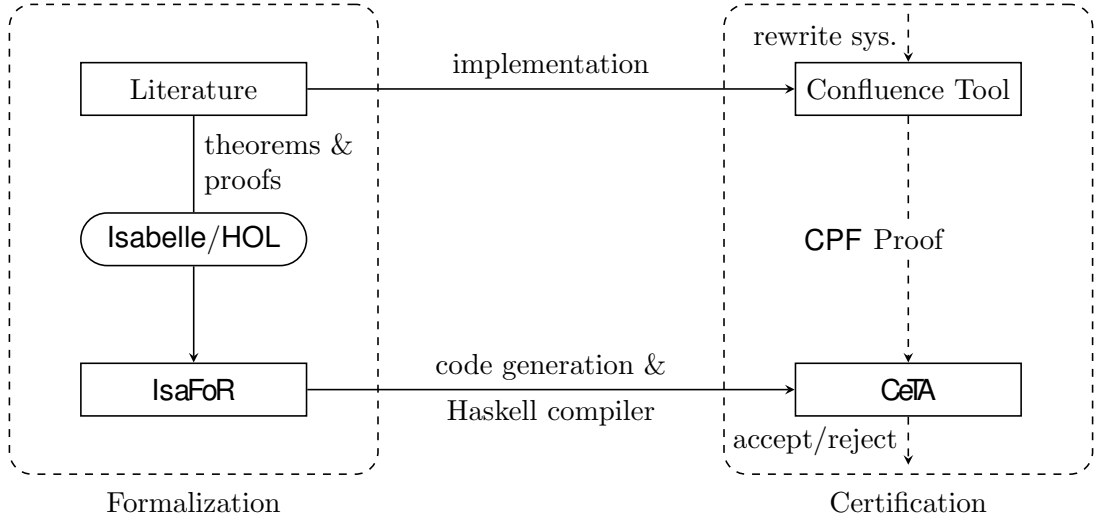

In this article we are concerned with the formalization and certification
of a recent confluence result in the proof assistant \ISABELLE/\HOL~\cite{ISABELLE}.
We based our formalization on the \textsf{Isa}belle \textsf{F}ormalization
\textsf{o}f \textsf{R}ewriting (\ISAFOR)~\cite{CETA},
adding approximately 4200 lines of
\ISABELLE code in \ISAR style.
\ISAFOR contains executable check functions
for each formalized proof technique together with formal proofs that whenever
such a check is accepted, the technique is applied correctly.
The certifier~\CETA is a trusted Haskell program obtained from
\ISAFOR, using \ISABELLE's code-generation facility~\cite{F09},
which ensures partial correctness~\cite{HN10}.
\CETA reads and checks proof
certificates in the certification problem format (\CPF)~\cite{CPF}.%
\footnote{\label{FOO:ceta}%
\ISAFOR/\CETA and \CPF are available at
  \url{http://cl-informatik.uibk.ac.at/software/ceta/}.}
The big picture of this automated, reliable confluence analysis is shown
in Figure~\ref{fig:certification},
the resources available from the URLs given in Footnotes~\ref{FOO:coco}
and \ref{FOO:ceta} provide more details.

\begin{figure}[t]
\centering{
{}
\hfill
\subfloat[\label{FIG:decreasing:valley}Valley version]{
\begin{tikzpicture}
\node (s) at (0,1) {$\cdot$};
\node (t) at (-3,0.5) {$\cdot$};
\node (u) at (3,0.5) {$\cdot$};
\node (t1) at (-2,-0.33) {$\cdot$};
\node (u1) at (2,-0.33) {$\cdot$};
\node (t2) at (-1,-1.16) {$\cdot$};
\node (u2) at (1,-1.16) {$\cdot$};
\node (v) at (-0,-2) {$\cdot$};
\draw[to] (s) to node [sloped,below] {$\scriptstyle\alpha$} (t);
\draw[to] (s) to node [sloped,below] {$\scriptstyle\beta$} (u);
\draw[to] (t) to
 node [sloped,below] {$\scriptstyle\Vee\alpha$} node [sloped,above] {$\scriptstyle *$} (t1);
\draw[to] (u) to
 node [sloped,below] {$\scriptstyle\Vee\beta$} node [sloped,above] {$\scriptstyle *$} (u1);
\draw[to] (t1) to
 node [sloped,below] {$\scriptstyle\beta$} node [sloped,above] {$\scriptstyle=$} (t2);
\draw[to] (u1) to
 node [sloped,below] {$\scriptstyle\alpha$} node [sloped,above] {$\scriptstyle=$} (u2);
\draw[to] (t2) to node [sloped,below] {$\scriptstyle\Vee\alpha\beta$} node [sloped,above] {$\scriptstyle*$} (v);
\draw[to] (u2) to node [sloped,below] {$\scriptstyle\Vee\alpha\beta$} node [sloped,above] {$\scriptstyle*$} (v);
\end{tikzpicture}
}
\hfill
\subfloat[\label{FIG:decreasing:conv}Conversion version]{
\begin{tikzpicture}
\node (s) at (0,1) {$\cdot$};
\node (t) at (-3.1,-1) {$\cdot$};
\node (u) at (3.1,-1) {$\cdot$};
\node (t1) at (-1.7,-1) {$\cdot$};
\node (u1) at (1.7,-1) {$\cdot$};
\node (t2) at (-0.7,-2) {$\cdot$};
\node (u2) at (0.7,-2) {$\cdot$};
\draw[to] (s) to node [sloped,below] {$\scriptstyle\alpha$} (t);
\draw[to] (s) to node [sloped,below] {$\scriptstyle\beta$} (u);
\draw[bi] (t) to
 node [sloped,below] {$\scriptstyle\Vee\alpha$} node [sloped,above] {$\scriptstyle *$} (t1);
\draw[bi] (u) to
 node [sloped,below] {$\scriptstyle\Vee\beta$} node [sloped,above] {$\scriptstyle *$} (u1);
\draw[to] (t1) to
 node [sloped,below] {$\scriptstyle\beta$} node [sloped,above] {$\scriptstyle=$} (t2);
\draw[to] (u1) to
 node [sloped,below] {$\scriptstyle\alpha$} node [sloped,above] {$\scriptstyle=$} (u2);
\draw[bi] (t2) to node [below] {$\scriptstyle\Vee\alpha\beta$} node [above] {$\scriptstyle*$} (u2);
\end{tikzpicture}
\hfill
{}
}
}
\caption{Locally decreasing diagrams.}
\label{FIG:decreasing}
\end{figure}
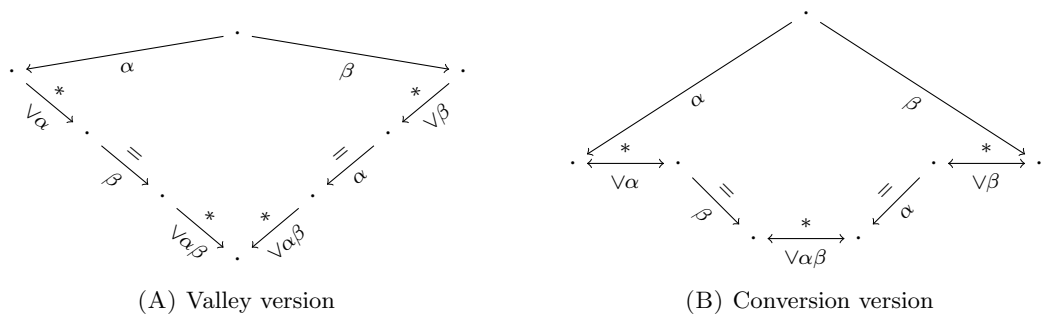

Decreasing diagrams~\cite{vO94} 
provide a complete characterization of confluence
for abstract rewrite systems whose convertibility classes are countable.
This confluence criterion states that a labeled abstract rewrite system
is confluent if each of its local peaks can be \emph{joined decreasingly}
in the shape of Figure~\ref{FIG:decreasing}\subref{FIG:decreasing:valley}, where $\Vee\alpha$
indicates that any label below $\alpha$ may be used, and $>$ is a
well-founded order on the labels. We will introduce these notions
formally in Section~\ref{PRE:main}.
As a criterion for abstract rewrite systems, decreasing diagrams can
be applied to first- and higher-order rewriting, including term rewriting and the 
$\lambda$-calculus. 

In this article we describe our formalization in {\ISABELLE}%
\footnote{In this article, the verb \emph{formalize} refers exclusively to
mechanized definitions and proofs in \ISABELLE/\HOL.}
of a recent powerful confluence result for term rewrite systems, exploiting
relative termination and decreasing diagrams.
Our aim was to formalize \cite[Corollary~16]{ZFM15} for a specific labeling
(which assigns labels to rewrite steps that are suitable for establishing
confluence by decreasing diagrams)
culminating in Theorem~\ref{THM:main}. Actually the formal proof of
Theorem~\ref{THM:main} is obtained by instantiating the more general result
of Theorem~\ref{THM:main:conv} appropriately.
Hence, confluence proofs according to these theorems are now checkable by~\CETA.
We achieved this via the following steps:
\begin{itemize}
\item Formalize local decreasingness for abstract rewrite systems
following~\cite{FvO13,F15d} (Section~\ref{ARS:main}).
\item Perform a detailed analysis of how local peaks can be joined for
(left-)linear term rewrite systems.
Note that this analysis has been performed in \ISABELLE and hence
\ISAFOR has been extended appropriately (Section~\ref{TRS:lp}).
\item Use the detailed analysis of local peaks to formalize the notion of a
labeling and a confluence result for term rewrite systems parametrized by a
labeling (Section~\ref{TRS:ld}).
In this way it is ensured that the formalization is reusable for other 
labelings stemming from~\cite{ZFM15}.
\item Instantiate the result from Section~\ref{TRS:ld} to obtain concrete
confluence results. We demonstrate how this instantiation can be done,
culminating in a formal proof of
Theorem~\ref{THM:main:conv} (Section~\ref{APP:main}).
\item Finally we made our formalization executable to check proof certificates:
we suitably extended \CPF to represent proofs according to Theorem~\ref{THM:main:conv}
and implemented dedicated check functions in our formalization, enabling \CETA
to inspect, i.e., certify, such confluence proofs (Section~\ref{APP:cert}).
\end{itemize}

\noindent To indicate which parts of the formalization are contributed by this
article, it is stated explicitly whenever existing results from \ISAFOR are (re)used.
More generally, all results displayed as definition, lemma, or theorem in
explicit \LaTeX{} environments have been formalized and are thus contributions
of this article and as such new additions to \ISAFOR.
In the presentation we use different levels of abstraction.
Code listings represent the formalization literally as \ISABELLE source code,
\LaTeX{} environments display the formalization as written in \ISABELLE (but
based on mathematical notation) and the intermediate text provides a high-level
description of the formalization in words.
The formalization constitutes our main contribution; it is not our goal
to advance the theory of rewriting in this work.

The remainder of this article is organized as follows.
Preliminaries are introduced in the next section.
In Section~\ref{ARS:main}
we describe the formalization of decreasing diagrams for abstract rewrite systems.
Based on the results for abstract rewriting and the notion of a labeling,
Section~\ref{TRS:main} formulates conditions that limit the attention to
critical peaks rather than arbitrary local peaks in the case of first-order
term rewriting.
Section~\ref{APP:main} instantiates these results with concrete labelings
to obtain corollaries that ensure confluence.
Section~\ref{EXP:main} presents an experimental evaluation, before
we conclude in Section~\ref{CON:main}.
The full formalization is available from the URL in Footnote~\ref{FOO:ceta}.

This article is an extended version of~\cite{NZ15}. Apart from improvements
in the presentation, new contributions
comprise the support for the \emph{conversion version} of decreasing diagrams,%
\footnote{The conversion version of decreasing diagrams
allows the use of conversions instead of forward rewrite sequences for parts
with small labels, which can reduce the size of the diagrams considerably.
This is illustrated in Figure~\ref{FIG:decreasing}\subref{FIG:decreasing:conv}. Note that
the left diagram is an instance of the right one.}
cf.~\cite{vO08a}, culminating in the new result of Theorem~\ref{THM:main:conv}.
The proof of Theorem~\ref{THM:main:conv} is more involved than the one of
Theorem~\ref{THM:main}, the target of~\cite{NZ15}, which is obtained here by
instantiating Theorem~\ref{THM:main:conv} appropriately.
Also the formalization described in Section~\ref{ARS:main} is new.
Compared to~\cite{NZ15}, the formalization of decreasing diagrams
in Section~\ref{ARS:main} uses an algebraic approach based on involutive
monoids, and is significantly smaller than the previous one.
The main motivation for the replacement was to cover decreasing diagrams
for commutation and Church-Rosser modulo, but those extensions are part
of ongoing \ISAFOR developments and are out of scope for this article.
However, as is common software engineering practice, we eliminated all
uses of the formalization presented in~\cite[Section~3]{NZ15},
thus reducing future maintenance effort in \ISAFOR.

\section{Preliminaries}
\label{PRE:main}
 
We assume familiarity with rewriting~\cite{BN98,TeReSe} and
decreasing diagrams~\cite{vO94}.
Basic knowledge of \ISABELLE~\cite{ISABELLE} is not essential but experience
with an interactive theorem prover might be helpful.

Let $\FF$ be a signature and $\VV$ a set of variables disjoint
from $\FF$. By $\FVTERMS$, we denote the set of \emph{terms} over $\FF$ and~$\VV$. 
Let~$\Box$ be a function symbol of arity zero and $\Box \notin \FF$.
A \emph{context} is a term in $(\FF \cup \{\Box\}, \VV)$, with a single
occurrence of~$\Box$, dubbed the \emph{context hole}.
\emph{Positions} are strings of positive natural numbers, i.e., elements of
$\Nat_+^*$.
We write $q \leqslant p$ if $qq' = p$ for some position $q'$,
in which case $p \backslash q$ is defined to be $q'$.
Furthermore $q < p$ if $q \leqslant p$ and $q \neq p$.
Finally, $q \parallel p$ if neither $q \leqslant p$ nor $p < q$.
Positions are used to address subterm occurrences.
The \emph{set of positions} of a term $t$ is defined as
$\Pos(t) = \{ \epsilon \}$ if $t$ is a variable and as
$\Pos(t) = \{ \epsilon \} \cup \{ iq \mid
\text{$1 \leqslant i \leqslant n$ and
$q \in \Pos(t_i)$} \}$ if $t = f(\seq{t})$.
The \emph{subterm} of $t$ at position $p \in \Pos(t)$ is defined as
$t|_p = t$ if $p = \epsilon$ and as
$t|_p = t_i|_q$ if $p = iq$ and $t = f(\seq{t})$. We write 
$s[t]_p$ for the result of replacing the occurrence of $s|_p$ at position~$p$
with $t$ in $s$.
The set of \emph{function symbol positions} $\FPos(t)$ is
$\{ p \in \Pos(t) \mid t|_p \notin \VV \}$
and $\VPos(t) = \Pos(t) \setminus \FPos(t)$.

A \emph{rewrite rule} is a pair of terms $(l,r)$, written $l \to r$.%
\footnote{We do not require the common \emph{variable conditions}, i.e., 
the restriction that $l$ is not a variable and all variables in $r$
are contained in $l$.}
A rewrite rule $l \to r$ is \emph{duplicating} if $|l|_x < |r|_x$ for some $x \in \VV$,
where the expression $|t|_x$ indicates the \emph{number of occurrences}
of the variable $x$ in term $t$.
A \emph{term rewrite system} (TRS) is a signature together with a 
set of rewrite rules over this signature. 
In the sequel, signatures are left implicit.
By $\RRd$ and $\RRnd$, we denote the \emph{duplicating} and \emph{non-duplicating
rules} of a TRS $\RR$, respectively.
A \emph{rewrite relation} is a binary relation on terms that is
closed under contexts and substitutions. For a TRS $\RR$ we define
$\to_\RR$ (often written as $\to$)
to be the smallest rewrite relation that contains~$\RR$.
As usual $\to^=$, $\to^+$, and $\to^*$ denote the reflexive,
transitive, and reflexive and transitive closure of $\to$, respectively,
while $\to^n$ denotes the $n$-fold composition of $\to$.
A \emph{relative TRS} $\RS$ is a pair of TRSs $\RR$ and $\SS$ with the
induced rewrite relation 
${\to_\RS} = {\to_{\SS}^* \cdot \to_{\RR}^{} \cdot \to_{\SS}^*}$,
where $\cdot$ denotes relation composition.
Sometimes we identify a TRS $\RR$ with the relative
TRS $\REL{\RR}{\varnothing}$ and vice versa,
which is justified by ${\to_{\RR/\varnothing}} = {\to_{\RR}}$.
A~TRS $\RR$ is \emph{terminating (relative to a TRS~$\SS$)} if $\to_\RR$
($\to_\RS$) is well-founded.
A TRS~$\RR$ is \emph{confluent} if
${\mathrel{{}_\RR^*{\from}} \cdot \to_\RR^*} \subseteq 
 {\to_\RR^* \cdot \mathrel{{}_\RR^*{\from}}}$.

A \emph{critical overlap}
$(l_1 \to r_1,p,l_2 \to r_2)_\mu$ of a TRS $\RR$ consists
of variants $l_1 \to r_1$ and $l_2 \to r_2$ of rewrite rules in $\RR$
without common variables, a position $p \in \FPos(l_2)$, and a most
general unifier $\mu$ of $l_1$ and $l_2|_p$.
From a critical overlap
$(l_1 \to r_1,p,l_2 \to r_2)_\mu$ we obtain a 
\emph{critical peak} $l_2\mu[r_1\mu]_p \cps{l_2\mu} r_2\mu$ and a
\emph{critical pair} $l_2\mu[r_1\mu]_p \cp r_2\mu$.%
\footnote{We based the definition of critical pairs on the one in \ISAFOR.
Note that \ISAFOR does not exclude root overlaps of a
rule with (a variant of) itself as is commonly done in the literature. This allows to
dispense with the variable condition that all variables in the
right-hand side of a rule must also occur on the left. Moreover, if a TRS
does satisfy the condition then all extra critical pairs that
would normally be excluded are trivial.}

If $l \to r \in \RR$, $p$ is a position, and $\sigma$ is a substitution
we call the triple $\pi = \langle p, l \to r, \sigma \rangle$ a
\emph{redex pattern}, and write $p_\pi$, $l_\pi$, $r_\pi$, $\sigma_\pi$ for its
position, left-hand side, right-hand side, and substitution, respectively.
We write $\xr{\pi}$ (or $\xr{p_\pi,l_\pi\to r_\pi, \sigma_\pi}$)
for a rewrite step at position $p_\pi$ using the rule
$l_\pi \to r_\pi$ and the substitution $\sigma_\pi$.
A redex pattern
$\pi$ \emph{matches} a term $t$ if $t|_{p_\pi} = l_\pi\sigma_\pi$,
in which case $t|_{p_\pi}$ is called a \emph{redex}.
Let $\pi_1$ and $\pi_2$ be redex patterns that match a common term.
They are called \emph{parallel}, written $\pi_1 \parallel \pi_2$,
if $p_{\pi_1} \parallel p_{\pi_2}$.
If $P = \{\pi_1,\pi_2,\ldots,\pi_n\}$ is a set of pairwise parallel
redex patterns matching a term
$t$, we denote by $t \xR{P} t'$ the \emph{parallel rewrite step} from $t$
to $t'$ by $P$, i.e., $t \to^{\pi_1} \cdot \to^{\pi_2} \cdot \ldots \cdot \to^{\pi_n} t'$.

An \emph{abstract rewrite system} (ARS) is a binary relation $\to$.
Let~$I$ be an index set.
We write $\{ \to_\alpha \}_{\alpha \in I}$ to
denote the ARS $\to$ where $\to$ is the union of
$\to_\alpha$ for all ${\alpha \in I}$. 
Let $\{ \to_\alpha \}_{\alpha \in I}$ be an ARS and
let $>$ and $\geqslant$ be relations on~$I$. 
Two relations $>$ and $\geqslant$ are called \emph{compatible} if 
${\geqslant} \cdot {>} \cdot {\geqslant} \subseteq {>}$.
Given a relation~$\rel$ we write 
$\xr[\Vrel\alpha_1 \cdots\,\alpha_n]{}$ for the union of all $\to_\beta$
with $\alpha_i \rel \beta$ for some $1 \leqslant i \leqslant n$.
Similarly, ${\Vrel S}$ is the set of all $\beta$ such that $\alpha \rel \beta$ 
for some $\alpha \in S$.
We call $\alpha$ and $\beta$ \emph{extended locally decreasing}
(for $>$ and~$\geqslant$) if
$
{\xl[\alpha]{}} \cdot {\xr[\beta]{}} \subseteq 
{\xc[\Vee\alpha]{*}} \cdot {\xr[\Veq\beta]{=}} \cdot
{\xc[\Vee\alpha\beta]{*}} \cdot
{\xl[\Veq\alpha]{=}} \cdot {\xc[\Vee\beta]{*}}
$.
If there exist a well-founded order~$>$ and a preorder $\geqslant$,
such that $>$ and $\geqslant$ are compatible, and
$\alpha$ and $\beta$ are extended locally decreasing
for all $\alpha, \beta \in I$ then the
ARS $\{ \to_\alpha \}_{\alpha \in I}$ is
\emph{extended locally decreasing} (for $>$ and $\geqslant$).
We call an ARS \emph{locally decreasing} (for~$>$) if it is
extended locally decreasing for $>$ and $=$, where the latter
is the identity relation.
In the sequel, we often refer to extended locally decreasing
as well as to locally decreasing just by decreasing, whenever the
context clarifies which concept is meant or the exact meaning
is irrelevant. In the literature the above condition is referred to 
as \emph{the conversion version of decreasing diagrams}. In its \emph{valley
version}, the condition reads as follows:
$
{\xl[\alpha]{}} \cdot {\xr[\beta]{}} \subseteq 
{\xr[\Vee\alpha]{*}} \cdot {\xr[\Veq\beta]{=}} \cdot
{\xr[\Vee\alpha\beta]{*}} \cdot {\xl[\Vee\alpha\beta]{*}} \cdot
{\xl[\Veq\alpha]{=}} \cdot {\xl[\Vee\beta]{*}}
$.
In the sequel we always refer to the conversion version of decreasing
diagrams, except when stated otherwise explicitly.

In \ISAFOR, the above notions are formalized essentially as described above.
However, since the underlying logic, \ISABELLE/\HOL, is typed, all notions are
parametrized by some base types.
For example, ARSs have an associated type for the objects being rewritten.
In the case of $\{\to_\alpha\}_{\alpha \in I}$,
there is one object type shared by all relations $\to_\alpha$ and
another type for the labels.
Terms and TRSs are parametric in the types of function symbols and variable symbols.
Fortunately, types have almost no impact on the proofs.
We will, therefore, not mention any types in the remainder of this
article.
\newpage

\section{Formalized Confluence Results for Abstract Rewriting}
\label{ARS:main}

This section is concerned with the formalization of the following
result from~\cite[Theorem~2]{HM10}:

\begin{lem}
\label{LEM:ars:eld}
Every extended locally decreasing ARS is confluent.
\qed
\end{lem}

It is interesting to note that in \cite{HM11}, the journal version of
\cite{HM10}, extended local decreasingness is avoided by employing the
\emph{predecessor labeling}.
Originally the authors used the \emph{source labeling}, wherein a
step $s \to t$ is labeled by $s$, and ${\geqslant} = {\to^*}$ for the
weak order. Consequently, any rewrite step $s' \to t'$ with $s \to^* s'$
has a label $s'$ with $s \geqslant s'$.
In the predecessor labeling, a rewrite step can be labeled by an
arbitrary predecessor, i.e., we have $s \to_u t$ if $u \to^* s$.
Now, if $s \to_u t$ and $s \to^* s' \to t'$, we may label the step from
$s'$ to $t'$ by $u$, the same label as $s \to_u t$.
In this way, the need for the weak comparison $\geqslant$, and hence
\emph{extended} local decreasingness, is avoided.
As the proof of Lemma~\ref{LEM:ars:eld} (see below) demonstrates,
this idea works in general, i.e.,
extended local decreasingness can be traded
for assigning more than one label to each rewrite step.
In the context of automation, however,
assigning just one (computable)
label to every rewrite step is very attractive,
as confluence tools must show critical peaks decreasing and confluence
certifiers must check the related proof certificates.

In the formalization, we first establish results for local decreasingness,
and then show that every extended locally decreasing ARS is also locally
decreasing. For the first part, we closely follow the development of
the \emph{monotonic} order in~\cite{FvO13}.
Note that the proof development in \cite{NZ15} was based
on~\cite{Z13}, which formalized the valley and conversion versions of
decreasing diagrams.
The formalization presented in this section goes beyond that, providing
results for commutation and for Church-Rosser modulo, see~\cite{FvO13}.
Despite covering more results, the development is still shorter than
the replaced one.
The key proof technique is that conversions are mapped to
\emph{French strings}, which abstract from the objects, keeping only
the labels and directions of the rewrite steps indicated by accents:
an acute accent ($\vl{\alpha}$) denotes a leftward step,
while a grave accent ($\vr{\alpha}$) marks a rightward step.
For example, the conversion
$
\m a \xl[\alpha]{} \m b \xr[\beta]{} \m c \xl[\alpha]{} \m b \xr[\gamma]{} \m d
$
would be represented by $\smash{\vl{\alpha}\vr{\beta}\vl{\alpha}\vr{\gamma}}$.
In fact we directly formalize \emph{Greek strings}, which additionally
allow macron accents ($\vu{\alpha}$) that may be used to represent
equational steps for results on Church-Rosser modulo
(cf.~\cite[Section 5]{FvO13}). This extension is
not used for proving Lemma~\ref{LEM:ars:eld}.
Greek strings are compared by a well-founded order that shows that
pasting local diagrams (which corresponds to replacing a substring
$\vl{\alpha}\vr{\beta}$ by a new string representing the joining
conversion in the decreasing diagram) terminates.
Because Greek strings are simply lists of Greek letters, we can
leverage existing results for lists from \ISABELLE/\HOL, which is one
reason why the formalization presented here is shorter.
In contrast, if one were to work directly on conversions, the proofs
would become cluttered by preconditions whenever two conversions are
concatenated, stating that the last object of the first conversion must
equal the initial object of the second conversion.
Another reason that the formalization is now shorter is that the new proofs are much more algebraic
(whereas diagram pasting as formalized in~\cite{Z13} has more of a geometric flavor),
and thus better suited for interactive theorem provers.

The full formalization of Lemma~\ref{LEM:ars:eld} is part of the Archive of Formal Proofs~\cite{F15d}.
There are few differences between the formalization and the paper version \cite{FvO13},
and they are mostly of technical nature.
For example, \cite[Definition~19]{FvO13}, which defines $\gg_\bullet$
(a well-founded order on Greek strings)
in terms of $>$ (a fixed well-founded order on labels),
is presented as a \emph{``proper recursive definition''} in the paper.
This idea is hard to convey to \ISABELLE
(it would require an elaborate termination proof),
so in the formalization, $\gg_\bullet$ is instead defined
by its characterization as the least fixed point of the map $\Lambda_{>}$
from the proof of \cite[Lemma~21]{FvO13},
(which states that $\gg_\bullet$ is a strict partial order).
\begin{listing}[t]
\Snippet{greek_less_def}
\caption{Definition of $(\gg_\bullet)^{-1} = \snippet{greek_less}\ (<)$}
\label{lst:greek_less}
\end{listing}
The resulting \ISABELLE definition is given in Listing~\ref{lst:greek_less}.
In a nutshell, $\snippet{ms_of_greek}$ maps a Greek string $x$
(which is a list of pairs of accents and labels)
to a corresponding multiset ($\langle x \rangle^g$ in~\cite{FvO13}),
using auxiliary functions $\snippet{adj_msog}$ and $\snippet{list_splits}$.
The latter returns all possible ways of splitting a list into a prefix,
a single element and a suffix, e.g.,
$\snippet{list_splits}\ [1,2,3] = [([\,],1,[2,3]),([1],2,[3]),([1,2],3,[\,])]$.
Next, $(\snippet{letter_less}\ (<))$ lifts $<$ to Greek letters,
and $(\snippet{nest}\ (<))$ corresponds to $\Lambda_{>}$,
employing the multiset extension $\snippet{mult}$ and a lexicographic product.
Finally, $(\snippet{greek_less}\ (<))$ computes $({\gg_\bullet})^{-1}$
as a least fixed point using $\snippet{lfp}$,
which is part of \ISABELLE/\HOL's library.

A notable difference between the paper and the formalization
concerns the proof of well-foundedness of the
order $\gg_\bullet$ on Greek strings, where instead of simple termination,
we employ Higman's lemma, which is more fundamental and was
already available in the Archive of Formal Proofs~\cite{S12,S14}.
\begin{thm}[{cf.\ \cite[Theorem 23]{FvO13}}]
If $>$ is a well-founded order on labels,
then $\gg_\bullet$ is a well-founded, monotonic,
partial order on Greek strings.
\end{thm}
\proof
We only show well-foundedness of $\gg_\bullet$ here.
Because $>$ is well-founded, it can be extended to a well-order $>'$.
Then $\geqslant'$ is a well-quasi-order.
When applied to Greek letters, $\geqslant'$ compares the underlying labels.
This lifted version is also a well-quasi-order.
In the corresponding order on Greek strings $\gg'_\bullet$ we have
$\vlr\alpha \gg'_\bullet \epsilon$ for all labels $\alpha$,
and $\vlr\alpha \gg'_\bullet \vlr\beta$ whenever $\vlr\alpha \mathrel{>}' \vlr\beta$.
By monotonicity~\cite[Lemma 22]{FvO13}, this implies
${\geqslant'_{\mathsf{emb}}} \subseteq ({\gg'_\bullet})^=$,
where $\geqslant'_{\mathsf{emb}}$ is the \emph{string embedding relation} induced by $\rel'$.
By Higman's lemma, $\geqslant'_{\mathsf{emb}}$ is a well-quasi-order,
and therefore $(\gg'_\bullet)^=$ is also a well-quasi-order.
The strict part of $(\gg'_\bullet)^=$ is $\gg'_\bullet$,
which is therefore a well-founded order,
and so is its subset $\gg_\bullet$.
\qed

In the rest of this section, we describe our formalized proof of
Lemma~\ref{LEM:ars:eld}.
Given an ARS that is extended locally decreasing for $>$ and $\geqslant$,
the proof in~\cite{HM10} constructs a single order $\succ$ on sets of
labels and establishes local decreasingness of the ARS for~$>$.
We do the same, but use a simplified proof for the following key lemma,
which allows us to use ${>}$ directly.

\begin{lem}
\label{LEM:ars:eld_imp_ld}
Every extended locally decreasing ARS is locally decreasing.
\end{lem}

The idea is to 
set ${\To_\alpha} := {\to_{\Veq\alpha}}$ and show that
$\To$ is locally decreasing.%
\footnote{Compared to \cite{HM10}, we use $\Veq \alpha$ instead of
$C_\alpha = (\Veq\alpha) \setminus (\Vee\alpha)$.}
This establishes the claim because
$\bigcup_{\alpha \in I} {\To_\alpha}
= \bigcup_{\alpha \in \Veq I} {\to_\alpha}
= \bigcup_{\alpha \in I} {\to_\alpha}$, and therefore,
${\To} = \{\To_\alpha\}_{\alpha\in I}$ and
${\to} = \{\to_\alpha\}_{\alpha \in I}$ are the same ARS,
labeled in two different ways.
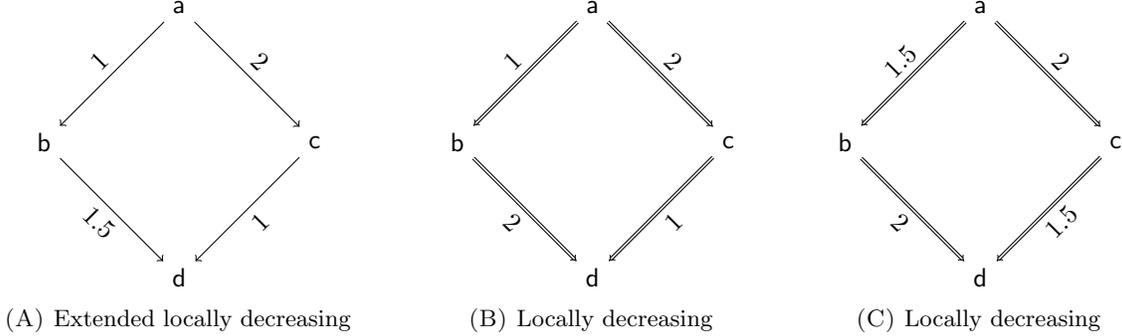
\begin{figure}
\subfloat[\label{FIG:EX:eld}Extended locally decreasing]{
\begin{tikzpicture}[scale=0.6]
\small
\node at (-0.7,3) {};
\node at (6.7,3) {};
\node at (3,6) (s)            {$\f{a}$};
\node at (0,3) (t)            {$\f{b}$};
\node at (6,3) (u)            {$\f{c}$};
\node at (3,0) (v)            {$\f{d}$};
\draw[to] (s) to node[anchor=south,sloped] {$1$} (t);
\draw[to] (t) to node[anchor=north,sloped] {$1.5$} (v);
\draw[to] (s) to node[anchor=south,sloped] {$2$} (u);
\draw[to] (u) to node[anchor=north,sloped] {$1$} (v);
\end{tikzpicture}
}
\hfill
\subfloat[\label{FIG:EX:ld1}Locally decreasing]{
\begin{tikzpicture}[scale=0.6]
\small
\node at (3,6) (s)            {$\f{a}$};
\node at (0,3) (t)            {$\f{b}$};
\node at (6,3) (u)            {$\f{c}$};
\node at (3,0) (v)            {$\f{d}$};
\draw[imp] (s) to node[anchor=south,sloped] {$1$} (t);
\draw[imp] (t) to node[anchor=north,sloped] {$2$} (v);
\draw[imp] (s) to node[anchor=south,sloped] {$2$} (u);
\draw[imp] (u) to node[anchor=north,sloped] {$1$} (v);
\end{tikzpicture}
}
\hfill
\subfloat[\label{FIG:EX:ld2}Locally decreasing]{
\begin{tikzpicture}[scale=0.6]
\small
\node at (3,6) (s)            {$\f{a}$};
\node at (0,3) (t)            {$\f{b}$};
\node at (6,3) (u)            {$\f{c}$};
\node at (3,0) (v)            {$\f{d}$};
\draw[imp] (s) to node[anchor=south,sloped] {$1.5$} (t);
\draw[imp] (t) to node[anchor=north,sloped] {$2$} (v);
\draw[imp] (s) to node[anchor=south,sloped] {$2$} (u);
\draw[imp] (u) to node[anchor=north,sloped] {$1.5$} (v);
\end{tikzpicture}
}
\caption{(Extended) locally decreasing peaks.}
\label{FIG:EX}
\end{figure}
The next example demonstrates some peculiarities of this approach.
\begin{exa}
\label{EX:ars}
Consider the ARS $\{\to_\alpha\}_{\alpha \in \{1,1.5,2\}}$ where
${\to_1} = \{(\f{a},\f{b}),(\f{c},\f{d})\}$, 
${\to_{1.5}} = \{(\f{b},\f{d})\}$, and
${\to_{2}} = \{(\f{a},\f{c})\}$. This ARS is extended locally decreasing for
the orders ${>} := \{(x,y) \mid x,y \in \Rat_{\geqslant 0}, x-y \geqslant 1\}$
and $\geqslant_\Rat$, as depicted
in Figure~\ref{FIG:EX}\subref{FIG:EX:eld}.
We have
${\Xr[2]{}} = \{(\f{a},\f{b}),(\f{c},\f{d}),(\f{a},\f{c}),(\f{b},\f{d})\}$,
${\Xr[1.5]{}} = \{(\f{b},\f{d}),(\f{a},\f{b}),(\f{c},\f{d})\}$, and
${\Xr[1]{}} = \{(\f{a},\f{b}),(\f{c},\f{d})\}$,
where e.g.\ ${\to_{1.5}} \subseteq {\Xr[2]{}}$ since ${2 \geqslant_\Rat 1.5}$.
Consequently, ${\To} = {\Xr[1]{}} \cup {\Xr[1.5]{}} \cup {\Xr[2]{}}$.
To establish local decreasingness
of the related ARS $\{\Xr[\alpha]{}\}_{\alpha\in \{1,1.5,2\}}$ the peak 
$\f{b} \Xl[1]{} \f{a} \Xr[2]{} \f{c}$ 
(emerging from $\f{b} \xl[1]{} \f{a} \xr[2]{} \f{c}$) must be considered,
which can be closed in a locally decreasing fashion
via $\f{b} \Xr[2]{} \f{d} \Xl[1]{} \f{c}$ 
(based on $\f{b} \xr[1.5]{} \f{d} \xl[1]{} \f{c}$),
as in Figure~\ref{FIG:EX}\subref{FIG:EX:ld1}.
Note that $\m{b} \Xr[1.5]{} \m{d} \Xl[1]{} \m{c}$ would not establish decreasingness
for~$>$.
However, the construction also admits the peak
$\f{b} \Xl[1.5]{} \f{a} \Xr[2]{} \f{c}$, for which there is no peak
$\f{b} \xl[1.5]{} \f{a} \xr[2]{} \f{c}$ in the original ARS, as 
it does not contain the step $\f{b} \xl[1.5]{} \f{a}$. 
Still, this peak can be closed locally decreasing
for~$>$, cf.\ Figure~\ref{FIG:EX}\subref{FIG:EX:ld2},
based on the steps in the diagram of Figure~\ref{FIG:EX}\subref{FIG:EX:eld}.
\end{exa}

\proof[Proof of Lemma~\ref{LEM:ars:eld_imp_ld}]
We assume the ARS $\{\to_\alpha\}_{\alpha \in I} $ is extended locally decreasing
for~$>$ and~$\geqslant$ and establish local decreasingness of the ARS 
$\{\To_\alpha\}_{\alpha \in I}$ for $>$ by showing
\begin{equation}
\label{EQ:ld}
{\Xl[\alpha]{}} \cdot {\Xr[\beta]{}} \subseteq 
 {\Xlr[\Vee \alpha]{*}} \cdot {\Xr[\beta]{=}} \cdot {\Xlr[\Vee \alpha\beta]{*}} \cdot
 {\Xl[\alpha]{=}} \cdot {\Xlr[\Vee \beta]{*}}
\end{equation}
for $\alpha,\beta \in I$. Assume that $x \Xl[\alpha]{} {\cdot} \Xr[\beta]{} y$.
By the definition of $\To$, there are $\alpha' \leqslant \alpha$ and
$\beta' \leqslant \beta$ such that $x \xl[\alpha']{} {\cdot} \xr[\beta']{} y$.
Because $\to$ is extended locally decreasing, this implies that
\[
x \xlr[\Vee\alpha']{*} {\cdot} \xr[\Veq\beta']{=} {\cdot} \xlr[\Vee\alpha'\beta']{*} {\cdot}
\xl[\Veq\alpha']{=} {\cdot} \xlr[\Vee\beta']{*} y
\]
Consider a label $\gamma \in \Vee\alpha'$, i.e., $\gamma < \alpha'$.
By compatibility, this implies $\gamma < \alpha$, i.e., $\gamma \in \Vee\alpha$.
Consequently, ${\xlr[\Vee\alpha']{}} \subseteq {\Xlr[\Vee\alpha]{}}$,
because $\leqslant$ is reflexive.
Similarly, if $\gamma \in \Veq\beta'$ then $\gamma \in \Veq\beta$,
because $\geqslant$ is transitive,
and hence ${\xr[\Veq\beta']{=}} \subseteq {\Xr[\beta]{=}}$.
Continuing in this fashion we obtain
\[
x \Xlr[\Vee\alpha]{*} {\cdot} \Xr[\beta]{=} {\cdot} \Xlr[\Vee\alpha\beta]{*} {\cdot}
\Xl[\alpha]{=} {\cdot} \Xlr[\Vee\beta]{*} y
\]
This establishes~\eqref{EQ:ld}.
Consequently, $\To$ is locally decreasing.
\qed

\section{Formalized Confluence Results for Term Rewriting}
\label{TRS:main}

This section builds upon the result for ARSs from the previous section to
prepare for confluence criteria for TRSs. 
First we recall the rule labeling~\cite{vO08a}, which maps each rewrite
step to a natural number based on the rewrite rule applied in the step.
\begin{defi}
\label{DEF:rl}
Let $i\colon\RR\to \Nat$ be an index mapping, which associates to every
rewrite rule a natural number.
The function $\ell^i(s\to^\pi t) = i (l_\pi \to r_\pi)$ is called \emph{rule labeling}.
Labels due to the rule labeling are compared by $>_\Nat$ and $\geqslant_\Nat$.
\end{defi}

Our aim is to formalize the following theorem,
which is one of the main results of~\cite{ZFM15}:
\begin{thm}[Valley version of decreasing diagrams]
\label{THM:main}
A left-linear TRS is confluent
if its duplicating rules terminate relative to its other rules
and all its critical peaks are decreasing for the rule labeling.
\end{thm}

To support other confluence results,
in the formalization we did not follow the easiest path, i.e.,
suit the definitions and lemmas directly to Theorem~\ref{THM:main}.
Rather, we adopted the approach from~\cite{ZFM15}, where all 
results are established via \emph{labeling functions} (satisfying some 
abstract properties). Apart from avoiding
a monolithic proof, this has the advantage that similar proofs need not be
repeated for different labeling functions. Instead it suffices to establish that
the concrete labeling functions satisfy some abstract conditions. 
In this approach decreasingness is established in three steps. The first step
comprises joinability results for local peaks (Section~\ref{TRS:lp}).
The second step (Section~\ref{TRS:ld}) formulates abstract conditions 
with the help of \emph{labeling functions} that admit a finite characterization
of decreasingness of local peaks. 
Finally, based on the previous two steps, the third step (Section~\ref{APP:main})
then obtains confluence results by instantiating the abstract labeling functions
with concrete ones, e.g.\ the rule labeling.
So only the third step needs to be adapted when formalizing new labeling
functions, as steps one and two are unaffected.
The content of this section is part of the \texttt{Decreasing\_Diagrams2.thy}
theory in \ISAFOR.

Compared to \cite{NZ15}, we formalized a conversion version of Theorem~\ref{THM:main},
which we will describe later (cf.~Theorem~\ref{THM:main:conv}).
This generalization affects Sections~\ref{TRS:ld} and~\ref{APP:main}.

\subsection{Local Peaks}
\label{TRS:lp}

As \ISAFOR already supported Knuth-Bendix' criterion (see~\cite{ST13}), it
contained results for joinability of local peaks and the critical 
pair theorem (the terms obtained by a local peak in a left-linear TRS are 
joinable or an instance of a critical pair).
However, large parts of the existing formalization could not
be reused directly as the established results lacked information required
for ensuring decreasingness. For instance, to obtain decreasingness for
the rule labeling 
in case of a variable peak, the rewrite rules employed in
the joining sequences are crucial, but the existing formalization only states
that such a local peak is joinable. On the other hand, the existing notion 
of critical pairs from \ISAFOR could be reused as the foundation for 
critical peaks. Since the computation of critical pairs
requires a formalized unification algorithm, extending \ISAFOR admitted
focusing on the tasks related to decreasingness.

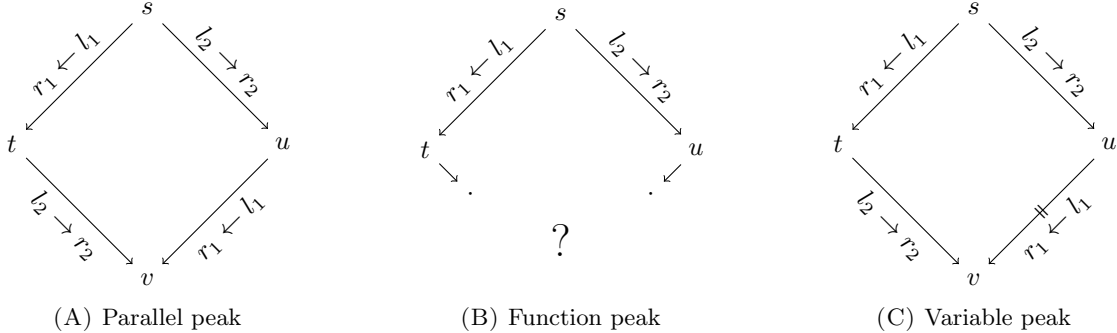
\begin{figure}[t]
\subfloat[\label{FIG:peaks:p}Parallel peak]{
\begin{tikzpicture}[scale=0.6]
\small
\node at (3,6) (s)            {$s$};
\node at (0,3) (t)            {$t$};
\node at (6,3) (u)            {$u$};
\node at (3,0) (v)            {$v$};
\draw[to] (s) to node[anchor=south,sloped] {$r_1\from l_1$} (t);
\draw[to] (t) to node[anchor=north,sloped] {$l_2\to r_2$} (v);
\draw[to] (s) to node[anchor=south,sloped] {$l_2\to r_2$} (u);
\draw[to] (u) to node[anchor=north,sloped] {$r_1\from l_1$} (v);
\end{tikzpicture}
}
\hfill
\subfloat[\label{FIG:peaks:o}Function peak]{
\begin{tikzpicture}[scale=0.6]
\small
\node at (3,6) (s)            {$s$};
\node at (0,3) (t)            {$t$};
\node at (6,3) (u)            {$u$};
\node at (1,2) (t1)           {$\cdot$};
\node at (5,2) (u1)           {$\cdot$};
\node at (3,0) (v)            {};
\node at (3,1)                {\LARGE ?};
\draw[to] (s) to node[anchor=south,sloped] {$r_1\from l_1$} (t);
\draw[to] (t) to node[anchor=north,sloped] {} (t1);
\draw[to] (s) to node[anchor=south,sloped] {$l_2\to r_2$} (u);
\draw[to] (u) to node[anchor=north,sloped] {} (u1);
\end{tikzpicture}
}
\hfill
\subfloat[\label{FIG:peaks:v}Variable peak]{
\begin{tikzpicture}[scale=0.6]
\small
\node at (3,6) (s)            {$s$};
\node at (0,3) (t)            {$t$};
\node at (6,3) (u)            {$u$};
\node at (3,0) (v)            {$v$};
\draw[to]   (s)  to node[anchor=south,sloped] {$r_1\from l_1$} (t);
\draw[to]   (t) to node[anchor=north,sloped] {$l_2\to r_2$} (v);
\draw[to]   (s)  to node[anchor=south,sloped] {$l_2\to r_2$} (u);
\draw[To]   (u)  to node[anchor=north,sloped] {$r_1\from l_1$} (v);
\end{tikzpicture}
}
\caption{Three kinds of local peaks.}
\label{FIG:peaks}
\end{figure}

Local peaks can be characterized based on the positions of the diverging
rewrite steps. Either the positions are parallel, called a
\emph{parallel peak}, or one
position is above the other. In the latter situation we further distinguish
whether the lower position is at a function position, 
called a \emph{function peak}, or
at/below a variable position of the other rule's left-hand
side, called a \emph{variable peak}. More precisely, for a local peak
\begin{equation}
\label{EQ:peak}
t = s[r_1\sigma_1]_p \from s[l_1\sigma_1]_p = s = s[l_2\sigma_2]_q \to
s[r_2\sigma_2]_q = u
\end{equation}
there are three possibilities (modulo symmetry):
\begin{description}
\item[\subref{FIG:peaks:p}]
$p \parallel q$
(parallel peak),
\item[\subref{FIG:peaks:o}]
$q \leqslant p$ and $p \backslash q \in \FPos(l_2)$
(function peak),
\item[\subref{FIG:peaks:v}]
\label{CASE:peaks:v}
$q \leqslant p$ and $p \backslash q \notin \FPos(l_2)$
(variable peak).
\end{description}

For the situation of a left-linear TRS these cases are visualized in
Figure~\ref{FIG:peaks}. 
It is easy to characterize parallel, function, and variable peaks in \ISABELLE
(cf.\ Listing~\ref{lst:peaks}) but it requires tedious notation.
E.g., the \snippet{local_peaks} of a TRS~$\RR$ are represented as the set of
all pairs $(s,r_1,p_1,\sigma_1,\snippet{True},t)$ and $(s,r_2,p_2,\sigma_2,\snippet{True},u)$,
with rewrite steps $s \to^{p_1,r_1,\sigma_1} t$ and $s \to^{p_2,r_2,\sigma_2} u$. 
The rewrite steps are formalized via the existing \ISAFOR notion of 
\snippet{rstep}, which represents a step $s \to^{p,l\to r,\sigma}_\RR t$
as \snippet{s_to_t}.
Here \snippet{True} and \snippet{False} are used to recall the direction of a
step, i.e., $s\to t$ and $t\xl[]{} s$, respectively.
This distinction is important for steps that are part of conversions,
because conversions may mix forward and backward steps.
In the definitions of \snippet{function_peak} and \snippet{variable_peak}
the symbol ${<}\#{>}$ is used, which is the \ISAFOR operation for concatenating
positions.

\begin{listing}[t!]
\Snippet{peaks}
\caption{Characterization of local peaks.}
\label{lst:peaks}
\end{listing}
\begin{listing}[t!]
\Snippet{peaks_cases}
\caption{Cases of local peaks.}
\label{lst:peaks_cases}
\end{listing}

As the definition of function and variable peaks is asymmetric the five cases of
local peaks can be reduced to the above three by mirroring those
peaks. 
Then local peaks can be characterized as in Listing~\ref{lst:peaks_cases}.
Next we elaborate on the three cases.

\paragraph*{Case 1: Parallel Peaks}

Figure~\ref{FIG:peaks}\subref{FIG:peaks:p} shows the shape of a local peak
where the steps take place at parallel positions. 
For a peak 
$
t \xl[]{\pi_1} s \xr[]{\pi_2} u
$
with $\pi_1 \parallel \pi_2$ we established that
$
t \xr[]{\pi_2} v \xl[]{\pi_1} u
$,
where opposing steps apply the same rule/substitution at the same position.
The proof is straightforward and based on a decomposition of the terms into
a context and the redex.

\paragraph*{Case 2: Function Peaks}

In general joining function peaks may involve rules not present in the
divergence (as indicated by the question mark in
Figure~\ref{FIG:peaks}\subref{FIG:peaks:o}). 
To reduce the burden of joining (infinitely many) function peaks
to joining the (in case of a finite TRS finitely many) critical peaks,
we formalized that every function peak is an instance of a critical peak.

\begin{lem}
\label{LEM:inst}
Let $t \xl{p, l_1 \to r_1, \sigma_1} s \xr{q, l_2 \to r_2, \sigma_2} u$ with
$qq' = p$, and $q' \in \FPos(l_2)$. Then there are a
context $C$, a substitution $\tau$, and a critical peak
$l_2\mu[r_1\mu]_{q'} \cps{l_2\mu} r_2\mu$ such that $s = C[l_2\mu\tau]$,
$t = C[(l_2\mu[r_1\mu]_{q'})\tau]$, and $u = C[r_2\mu\tau]$.
\qed
\end{lem}
This fact is folklore, see e.g.~\cite[Lemma 2.7.12]{TeReSe}.
We remark that this fact was already present (multiple times) in \ISAFOR,
but concealed in larger proofs, e.g.\ the formalization of
orthogonality~\cite{NT14}, and never stated explicitly.

As \ISAFOR does not enforce that the variables of a rewrite rule's right-hand
side are contained in its left-hand side, such rules are just also included in
the critical peak computation.

\paragraph*{Case 3: Variable Peaks}

Variable overlaps (Figure~\ref{FIG:peaks}\subref{FIG:peaks:v})
can again be joined by the rules involved in the diverging step.%
\footnote{This includes rules having a variable as left-hand side.}
We only consider the case if $l_2\to r_2$ is left-linear, as our main result
assumes left-linearity. 
More precisely, if $q'$ is the unique position in $\VPos(l_2)$ such
that $qq' \leqslant p$, $x = l_2|_{q'}$, and $|r_2|_x = n$ then
we have $t \xr[l_2 \to r_2]{} v$, which is similar to the case for parallel peaks, as the
redex $l_2\sigma$ becomes $l_2\tau$ but is not destroyed,
and $u \xr[l_1\to r_1]{n} v$.
To obtain this result we reason via parallel rewriting.
The notion of parallel rewriting already supported by \ISAFOR (employed to prove
that orthogonal systems are confluent) does not keep track of, for example, the applied
rules. Thus we augmented \ISAFOR by a new version of parallel steps, which
record the information (position, rewrite rule, substitution) of each rewrite
step, i.e., the rewrite relation is decorated with the contracted redex patterns.

\begin{defi}
For a TRS~$\RR$ the \emph{parallel rewrite relation}~$\xR[]{}$ is defined by the following
inference rules.
\begin{mathpar}
  \inferrule{ }{x \xR[]{\varnothing} x}
  \and
  \inferrule{l \to r \in \RR}
  {l\sigma \xR[]{\{\langle \epsilon ,l \to r,\sigma\rangle\}} r\sigma}
  \and
  \inferrule{s_1 \xR[]{P_1} t_1 \\ \cdots
    \\ s_n \xR[]{P_n} t_n}
  {f(s_1,\ldots,s_n) \xR[]{(1P_1)\cup \dotsb \cup (nP_n)} f(t_1,\ldots,t_n)}
\end{mathpar}
Here for a set of redex patterns $P = \{\pi_1,\ldots,\pi_m\}$ by $iP$ we denote
$\{i\pi_1,\ldots,i\pi_m\}$ with $i\pi = \langle ip, l \to r, \sigma \rangle$
for $\pi = \langle p, l \to r, \sigma \rangle$.
\end{defi}
To use this parallel rewrite relation for closing variable peaks we formalized
the following auxiliary facts.%
\footnote{In a typical mathematical proof, these facts would be considered as
self-evident. However, \ISABELLE is not so easily convinced, so a proof is
required. So rather than being an interesting result, Lemma~\ref{LEM:par}
serves to illustrate the level of detail that formalizations often require.}

\begin{lem}
\label{LEM:par}
The following properties of the parallel rewrite relation hold:
\begin{enumerate}
\item For all terms $s$ we have $s \xR[]{\varnothing} s$.
\item If $s \xR[]{\varnothing} t$ then $s = t$.
\item If $s \xR[]{P} t$ and $q \in \Pos(u)$ then $u[s]_q \xR[]{qP} u[t]_q$.
\item We have $s \to^\pi t$ if and only if $s \xR[]{\{\pi\}} t$.
\item If $\sigma(x) \to^\pi \tau(x)$ and $\sigma(y) = \tau(y)$ for all
  $y \in \VV$ with $y \neq x$ then $t\sigma \xR[]{P} t\tau$ with
  $l_{\pi'} \to r_{\pi'} = l_\pi \to r_\pi$ for all $\pi' \in P$.
\item If $s \xR[]{\{\pi\} \cup P} t$ then there is a term $u$ with
  $s \xR[]{\{\pi\}} u \xR[]{P} t$.
\item If $s \xR[]{\{\pi_1,\ldots,\pi_n\}} t$ then 
  $s \to^{\pi_1} \cdots \to^{\pi_n} t$.
\qed
\end{enumerate}
\end{lem}
In principle the statements of Lemma~\ref{LEM:par} follow from the definitions using straightforward
induction proofs, building upon each other in the order they are listed.
However, the additional bookkeeping, required to correctly
propagate the information attached to the rewrite relation, makes them
considerably more involved than for the existing, agnostic notion of parallel
rewriting.

Now for reasoning about variable peaks as in 
case \subref{FIG:peaks:v} on page~\pageref{CASE:peaks:v}
we decompose
$u = s[r_2\sigma]_q$ and $v=s[r_2\tau]_q$ where $\sigma(y) = \tau(y)$ for all
$y \in \VV \setminus \{x\}$ and
$\sigma(x) \to_{p \backslash qq', l_1\to r_1} \tau(x)$. From the latter by 
item~(5) we obtain $r_2\sigma \xR[]{P} r_2\tau$, where all redex patterns in $P$ use
$l_1 \to r_1$. Then by item~(3) we get $s[r_2\sigma]_q \xR[]{qP} s[r_2\tau]_q$
and finally $s[r_2\sigma]_q \xr[l_1\to r_1]{n} s[r_2\tau]_q$ with $n = |qP| = |P|$ by
item~(7).

\subsection{Local Decreasingness}
\label{TRS:ld}

This section presents a confluence result (cf.\ Corollary~\ref{COR:main})
based on decreasingness of the critical peaks. Abstract conditions,
via the key notion of a labeling, will ensure that parallel peaks
and variable peaks are decreasing. Furthermore, these conditions imply
that decreasingness of the critical peaks implies decreasingness of the
function peaks.
\begin{listing}[t]
\Snippet{conv}
\caption{Conversions.}
\label{lst:conv}
\end{listing}
For establishing (extended) local decreasingness, a label must be attached to 
rewrite steps. To facilitate the computation of labels,
in the formalization conversions provide
information about the intermediate terms, applied rules, etc.
In Listing~\ref{lst:conv} the definition of conversions as an inductive set
is provided. The first case states that the (empty) conversion starting from a 
term $s$ is a conversion. The second case states that if
$s \to_\RR t$ (using rule~$r$ at position~$p$ with substitution~$\sigma$) and
there is a conversion starting from~$t$ (where the list~$ts$ collects
the details on the conversion), then there also is a conversion starting
from~$s$ and the details for this conversion
are collected in the list where the tuple $(s,r,p,\sigma,\snippet{True},t)$
is added to the list~$ts$. While the first two cases of Listing~\ref{lst:conv}
amount to an inductive definition of rewrite sequences, together with the third
case (which considers a step $t \from s$ and a conversion starting from~$s$),
conversions are obtained.
Here \snippet{True} and \snippet{False} are used to recall the direction of a
step, as in Listing~\ref{lst:peaks}.

Furthermore, labels are computed by a \emph{labeling (function)}, having
(local) information about the rewrite step (such as source and target term,
applied rewrite rule, position, and substitution) it is expected to label.
For reasons of readability in this presentation we employ the 
mathematical notation (e.g., $\to^*$
for sequences and $\leftrightarrow^*$ for conversions,
etc.) with all information implicit
but remark that the formalization works on 
sequences and
conversions with explicit
information (as in Listing~\ref{lst:conv}).

\begin{defi}
\label{DEF:lab}
A \emph{labeling} is a function $\ell$ from rewrite steps to a set of labels such
that for all contexts $C$ and substitutions $\sigma$ the following properties
are satisfied:
\begin{itemize}
\item If $\ell(s \to^{\pi_1} t) > \ell( u \to^{\pi_2} v)$ then 
 $\ell(C[s\sigma] \to^{C[\pi_1\sigma]} C[t\sigma]) > \ell(C[u\sigma] \to^{C[\pi_2\sigma]} C[v\sigma])$
\item If $\ell(s \to^{\pi_1} t) \geqslant \ell( u \to^{\pi_2} v)$ then
 $\ell(C[s\sigma] \to^{C[\pi_1\sigma]} C[t\sigma]) \geqslant \ell(C[u\sigma] \to^{C[\pi_2\sigma]} C[v\sigma])$
\item $\ell(s \to^\pi t) = \ell(t \xl[]{\pi} s)$
\end{itemize}
Here $C[\pi\sigma]$ denotes $\langle qp, l \to r, \tau\sigma \rangle$ for
$\pi = \langle p, l \to r, \tau \rangle$ and $C|_q = \Box$.
\end{defi}

As the co-domain of a labeling is a set of labels, a labeling according to
Definition~\ref{DEF:lab} maps a single rewrite step to a single label, which is
different from how (some) ARSs are labeled in Section~\ref{ARS:main}.

\begin{exa}
The rule labeling is a labeling in our sense.
\end{exa}

In presence of a labeling, conversions can be labeled at any
time. This avoids lifting many notions (such as rewrite steps, local peaks, 
rewrite sequences, etc.) and results from rewriting to labeled rewriting.

In the next definition a labeling is applied to conversions
via the equations
${\ell(t \conv^0 t) = \varnothing}$,
${\ell(s\to^\pi t \conv^* u) = \{\ell(s\to^\pi t)\} \cup \ell(t \conv^* u)}$,
and ${\ell(s\xl{\pi} t \conv^* u) = \{\ell(s\xl[]{\pi} t)\} \cup \ell(t \conv^* u)}$.

\begin{defi}
A local peak $t \xl{\pi_1} s \xr{\pi_2} u$ is \emph{extended locally decreasing}
(for a labeling $\ell$ and orders~$>$ and~$\geqslant$) if there is
a local diagram, i.e.,
$t \xc{*} t' \xr{=} t'' \xc{*} u'' \xl{=} u' \xc{*} u$ such that
its labels are extended locally decreasing, i.e., 
\begin{itemize}
\item
$\ell(t\xc{*}t') \subseteq \Vee \ell(t \xl{\pi_1} s)$, 
\item
$\ell(t'\xr{=}t'') \subseteq \Veq \ell(s \xr{\pi_2} u)$, 
\item
$\ell(t''\xc{*}u'') \subseteq \Vee \ell(t \xl{\pi_1} s \xr{\pi_2} u)$,
\item
$\ell(u''\xl{=}u') \subseteq \Veq \ell(t \xl{\pi_1} s)$, and
\item
$\ell(u'\xc{*}u) \subseteq \Vee \ell(s \xr{\pi_2} u)$.
\end{itemize}
\end{defi}

Following~\cite{Z13}, we separate (local) diagrams (where rewriting is involved) 
from decreasingness (where only the labels are involved).
In contrast to the valley version of decreasing diagrams, as employed in~\cite{NZ15},
where a sequence $\to^*$ can be decomposed into
$\xr[\Vee\alpha]{*} \cdot \xr[\Veq\beta]{=} \cdot \xr[\Vee\alpha\beta]{*}$ based on 
the sequence of labels, this is no longer possible for the conversion version (as there is
no guarantee that the step $\xr[\Veq\beta]{=}$ is oriented correctly).
Hence we made the decomposition explicit for the conversion version.

The corresponding predicate in \ISAFOR is given in Listing~\ref{lst:eld} where 
extended local decreasingness (\snippet{eldc}) of a local peak \isa{p} is expressed
via the existence of conversions \snippet{cl1}, \snippet{cl2}, \snippet{cr1}, \snippet{cr2}
and possibly empty steps \snippet{sl} and \snippet{sr} that close the
divergence caused by the local peak \isa{p} in the shape of a local diagram
(\snippet{ldc_trs}).\footnote{The conversions \snippet{cl2} and \snippet{cr2} could be merged
into a single conversion. However, in proofs the symmetric version permits to reason about one
side and obtain the other side for free.}
Here \snippet{get_target} gets the target of a rewrite step and
\snippet{first} and \snippet{last} get the first and last element of a
conversion or rewrite sequence, respectively.
Moreover the labels of the underlying conversions are required to
be extended locally decreasing (\snippet{eld_conv}, \snippet{ELD_1}).
Here \snippet{ds} $(\reli)$ $S$ is the notation for ${\Vrel S}$ and
\isa{r} is the pair of relations $(>,\geqslant)$.

\begin{listing}[t]
\Snippet{eldc_def}
\Snippet{ldc_trs_def}
\Snippet{eld_conv_def}
\Snippet{ELD_1_def}
\caption{Extended local decreasingness.}
\label{lst:eld}
\end{listing}

Then a function peak is extended locally decreasing if the critical peaks are.
\begin{lem}
\label{LEM:main_eld}
Let $\ell$ be a labeling and let all critical peaks of a TRS~\RR 
be extended locally decreasing for $\ell$. Then every function peak of~\RR
is extended locally decreasing for~$\ell$.
\end{lem}
\proof
As every function peak is an instance of a critical peak (see Lemma~\ref{LEM:inst}),
the result follows from $\ell$ being a labeling (Definition~\ref{DEF:lab}).
\qed

The notion of compatibility (between a TRS and a labeling) admits a finite
characterization of extended local decreasingness.

\begin{defi}
\label{DEF:compatible}
Let $\ell$ be a labeling. We call $\ell$ \emph{compatible} with a TRS~$\RR$
if all parallel
peaks and all variable peaks of~\RR are extended locally decreasing for~$\ell$.

\end{defi}

The key lemma then establishes that if $\ell$ is compatible with a TRS,
then all local peaks are extended locally decreasing.

\begin{lem}
\label{LEM:main}
Let $\ell$ be a labeling that is compatible with a TRS~$\RR$. If the
critical peaks of~$\RR$ are extended locally decreasing for~$\ell$, then all local
peaks of~\RR are extended locally decreasing
for~$\ell$.
\end{lem}
\proof
The cases of variable and parallel peaks are taken care of by compatibility.
The case of function peaks follows from the assumption in connection with
Lemma~\ref{LEM:main_eld}. The symmetric cases for function and variable
peaks can be resolved by mirroring the local diagrams.
\qed

Representing a TRS $\RR$ over the signature~$\FF$ and variables~$\VV$ as the ARS
over objects $\FVTERMS$ and relations 
${\to_\alpha} = \{(s,t) \mid \text{$s \to^\pi t$ and $\ell(s \to^\pi t) = \alpha$}\}$
for some labeling~$\ell$,
Lemma~\ref{LEM:ars:eld} immediately applies to TRSs. To this end, extended local
decreasingness formulated via explicit rewrite sequences (with labeling functions)
has to be mapped to extended local decreasingness on families of (abstract rewrite)
relations; we omit the technical details here.
Finally, we are ready to formalize
a confluence result for first-order term rewriting.
\begin{cor}
\label{COR:main}
Let $\ell$ be a labeling compatible with a TRS~$\RR$,
and let $>$ and $\geqslant$ be a well-founded order and a compatible preorder, respectively.
If the critical peaks
of~\RR are extended locally decreasing for~$\ell$ and $>$ and $\geqslant$
then $\RR$ is confluent.
\qed
\end{cor}

Concrete confluence criteria are then obtained
as instances of the above result by instantiating~$\ell$.
For example, Theorem~\ref{THM:main} is obtained
by showing that its relative termination
assumption in combination with the rule labeling 
yields a compatible labeling for left-linear TRSs.

\section{Checkable Confluence Proofs}
\label{APP:main}

In this section we instantiate Corollary~\ref{COR:main} 
to obtain concrete confluence results; first for linear TRSs (Section~\ref{APP:l})
and then for left-linear TRSs (Section~\ref{APP:ll}).
Afterwards we discuss the design of the certificates, checkable by \CETA,
in Section~\ref{APP:cert}.
The first two subsections are part of \texttt{Decreasing\_Diagrams2.thy} in \ISAFOR,
whereas the executable versions from Section~\ref{APP:cert} can be found
in \texttt{Rule\_Labeling\_Impl.thy}.

\subsection{Checkable Confluence Proofs for Linear TRSs}
\label{APP:l}

The rule labeling admits a confluence criterion for linear TRSs
based on the formalization established so far.
\begin{lem}
\label{LEM:rl}
\hfill
\begin{enumerate}
\item \label{LEM:rl:1} The rule labeling is a labeling.
\item \label{LEM:rl:2} Parallel peaks are extended locally decreasing for the rule labeling.
\item \label{LEM:rl:3} Variable peaks of a linear TRS are extended locally decreasing for the rule labeling.
\item \label{LEM:rl:4} The rule labeling is compatible with a linear TRS.
\item \label{LEM:rl:5}
 A linear TRS is confluent if its critical peaks are extended locally decreasing for the rule labeling.
\end{enumerate}
\end{lem}
\proof
Item~\eqref{LEM:rl:1} follows from Definition~\ref{DEF:lab}
and Definition~\ref{DEF:rl}.
For items~\eqref{LEM:rl:2} and~\eqref{LEM:rl:3} we employ the analysis of
parallel and variable peaks from Section~\ref{TRS:lp}, respectively.
Item~\eqref{LEM:rl:4} is then a consequence of items~\eqref{LEM:rl:2} 
and~\eqref{LEM:rl:3}. Finally, item~\eqref{LEM:rl:5} amounts to an application
of Corollary~\ref{COR:main}.
\qed

We demonstrate the rule labeling on a simple example.
\begin{exa}
\label{ex:rl}
Consider the TRS consisting of the following rules,
where subscripts indicate labels for the rule labeling:
\begin{xalignat*}{5}
\m{a} \to_1 \m{b}&&
\m{a} \to_1 \m{d}&&
\m{b} \to_0 \m{a}&&
\m{c} \to_0 \m{a}&&
\m{c} \to_0 \m{b}
\end{xalignat*}
The critical peaks are decreasing for the rule labeling,
as depicted in Figure~\ref{fig:ex:rl}.
\end{exa}
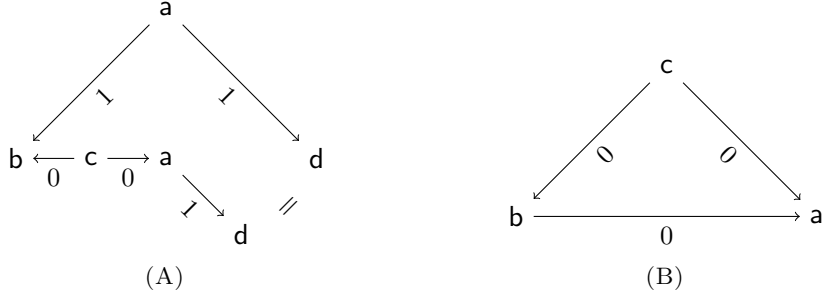
\begin{figure}
\hfill
\subfloat[\label{fig:ex:rl:1}]{
\begin{tikzpicture}[scale=1.0]
\node at (0,2)   (t)            {\m{a}};
\node at (-2,0)  (l1)           {\m{b}};
\node at (-1,0)  (l2)           {\m{c}};
\node at (0,0)   (l3)           {\m{a}};
\node at (1,-1)  (l4)           {\m{d}};
\node at (2,0)   (r1)           {\m{d}};
\draw[to] (t) to node[anchor=north,sloped] {\small 1} (l1);
\draw[to] (l2) to node[anchor=north,sloped] {\small 0} (l1);
\draw[to] (l2) to node[anchor=north,sloped] {\small 0} (l3);
\draw[to] (l3) to node[anchor=north,sloped] {\small 1} (l4);
\draw[to] (t) to node[anchor=north,sloped] {\small 1} (r1);
\draw[white] (r1) to node[anchor=north,sloped] {\textcolor{black}{$=$}} (l4);
\end{tikzpicture}
}
\hfill
\subfloat[\label{fig:ex:rl:2}]{
\begin{tikzpicture}[scale=2.0]
\node at (0,1)   (t)            {\m{c}};
\node at (-1,0)  (l1)           {\m{b}};
\node at (1,0)   (r1)           {\m{a}};
\draw[to] (t) to node[anchor=north,sloped] {\small 0} (l1);
\draw[to] (t) to node[anchor=north,sloped] {\small 0} (r1);
\draw[to] (l1) to node[anchor=north,sloped] {\small 0} (r1);
\end{tikzpicture}
}
\hfill{}
\caption{Decreasingness of critical peaks in Example~\ref{ex:rl}.}
\label{fig:ex:rl}
\end{figure}

Eventually, we remark that for the rule labeling extended local decreasingness
directly implies local decreasingness, as $\geqslant_\Nat$ is the reflexive closure
of $>_\Nat$. 

\subsection{Checkable Confluence Proofs for Left-linear TRSs}
\label{APP:ll}

That a locally confluent terminating left-linear TRS is confluent can be
established in the flavor of Lemma~\ref{LEM:rl}. The restriction to
left-linearity arises from the lack of considering non-left-linear variable
peaks in Section~\ref{TRS:lp}. As the analysis of such a peak would not give
further insights we pursue another aim in this section, i.e., the mechanized
proof of Theorem~\ref{THM:main}. 

It is well known that
the rule labeling $\ell^i$ is in general not compatible with
left-linear 
TRSs, cf.~\cite{HM11}.
Thus, to obtain extended local decreasingness for variable peaks,
in Theorem~\ref{THM:main}
the additional relative termination assumption is exploited. To this
end, we use the source labeling.

\begin{defi}
\label{DEF:sl}
The \emph{source labeling} maps a rewrite step to its source, i.e.,
${\ell^\src(s \to^\pi t) = s}$. Labels due to the source
labeling are compared by the orders $\to_{\RRdnd}^+$ and $\to^*_\RR$.
\end{defi}

The relative termination assumption of Theorem~\ref{THM:main} makes all
variable peaks of a left-linear TRS extended locally decreasing for the
source labeling.
These variable peaks might not be extended locally decreasing
for the rule labeling, as the step $u \xR[\{l_1\to r_1\}]{} v$ in
Figure~\ref{FIG:peaks}\subref{FIG:peaks:v} yields $u \to^n v$
for~$n$ possibly larger than one.
Hence we introduce weaker versions of decreasingness and compatibility.

\begin{defi}
A diagram of the shape
$t \xl[\alpha]{{}} s \xr[\beta]{l_2 \to r_2} u$,
$t \xr[\Veq \beta]{{}} v \xl[\Veq \alpha]{n} u$
is called \emph{weakly extended locally decreasing} if $n \leqslant 1$
whenever $r_2$ is linear.
We call a labeling $\ell$ \emph{weakly compatible} with a TRS~\RR
if parallel and variable peaks are weakly extended locally decreasing for $\ell$.
\end{defi}

Following~\cite{ZFM15}, the aim is to establish that the lexicographic
combination of a compatible labeling with a weakly compatible labeling
(e.g., $\ell^\src \times \ell^i$) is compatible with a left-linear TRS.
While weak extended local decreasingness could also be defined in the spirit of
extended local decreasingness (with a more complicated join sequence or
involving conversions), the chosen formulation suffices to establish the
result, eases the definition, and simplifies proofs. As this notion is
only used to reason about parallel and variable peaks, which need not be
processed by automatic tools, the increased generality would not
benefit automation.

Based on the peak analysis of Section~\ref{TRS:lp}, the following
results are formalized (such properties must be proved for each labeling function):

\begin{lem}
\label{LEM:rl2}
Let $\RR$ be a left-linear TRS.
\begin{enumerate}
\item
\label{LEM:rl2:pp}
Parallel peaks are weakly extended locally decreasing for the rule labeling.
\item
\label{LEM:rl2:vp}
Variable peaks of~\RR are weakly extended locally decreasing for the rule labeling.
\item \label{LEM:rl2:4}
The rule labeling is weakly compatible with~\RR.
\end{enumerate}
\end{lem}
\proof
Results~\eqref{LEM:rl2:pp} and~\eqref{LEM:rl2:vp} are established by labeling the rewrite
steps in the corresponding diagrams based on the characterizations of parallel and variable peaks
of Figures~\ref{FIG:peaks}\subref{FIG:peaks:p} and~\ref{FIG:peaks}\subref{FIG:peaks:v}, respectively.
Item~\eqref{LEM:rl2:4} follows from~\eqref{LEM:rl2:pp} and~\eqref{LEM:rl2:vp} together with
Lemma~\ref{LEM:rl}\eqref{LEM:rl:1}.
\qed

Similar results are formalized for the source labeling.

\begin{lem}
\label{LEM:sn}
Let $\RR$ be a left-linear TRS whose duplicating rules terminate relative to the other rules.
\begin{enumerate}
\item \label{LEM:sn:l}
The source labeling is a labeling.
\item \label{LEM:sn:p}
Parallel peaks are extended locally decreasing for the source labeling.
\item \label{LEM:sn:v}
Variable peaks of~\RR are extended locally decreasing for the source labeling.
\item \label{LEM:sn:3}
The source labeling is compatible with~\RR.
\end{enumerate}
\end{lem}
\proof
As rewriting is closed under contexts and substitutions, item~\eqref{LEM:sn:l} holds.
Items~\eqref{LEM:sn:p} and~\eqref{LEM:sn:v} are established
along the lines of the proof of Lemma~\ref{LEM:rl2}.
Finally, item~\eqref{LEM:sn:3} follows from~\eqref{LEM:sn:p} and~\eqref{LEM:sn:v} in 
combination with Definition~\ref{DEF:compatible}.
\qed

Using this lemma, we proved the following results for the lexicographic
combination of the source labeling with another labeling.

\begin{lem}
\label{LEM:lex}
Let $\RR$ be a left-linear TRS whose duplicating rules terminate relative to
the other rules and $\ell$ be a labeling weakly compatible with~\RR.
\begin{enumerate}
\item Then $\ell^\src \times \ell$ is a labeling.
\item \label{LEM:lex:2}
Then $\ell^\src \times \ell$ is compatible with~\RR.
\qed
\end{enumerate}
\end{lem}

For reasons of readability we have left the orders $>$ and $\geqslant$ that are
required for (weak) compatibility implicit and just mention that the lexicographic
extension (as detailed in~\cite{ZFM15}) preserves the required properties.
Finally, we prove Theorem~\ref{THM:main}.

\proof[Proof of Theorem~\ref{THM:main}]
\label{THM:main:proof}
From Lemma~\ref{LEM:rl2}(\ref{LEM:rl2:4}) in combination with
Lemma~\ref{LEM:lex} we obtain that $\ell^\src \times \ell^i$ is a labeling compatible with
a left-linear TRS, provided the relative termination assumption is satisfied.
By assumption, the critical peaks are (extended locally) decreasing for the rule
labeling~$\ell^i$. As along a rewrite sequence labels with respect to $\ell^\src$ never
increase, the critical peaks are extended locally decreasing for $\ell^\src \times \ell^i$.
We conclude the proof by an application of Corollary~\ref{COR:main}.
\qed

Actually a stronger result than Theorem~\ref{THM:main} has been
established, as $\ell^\src \times \ell^i$ might show more critical peaks decreasing
than $\ell^i$ alone.

\begin{figure}
\begin{tikzpicture}[scale=0.5]
\small
\node at (0,10)  (u)            {};
\node at (-10,0) (l1)           {};
\node at (-9,-1) (l2)           {};
\node at (-8,0)  (l3)           {};
\node at (-7,-1) (l4)           {};
\node at (-6,0)  (l5)           {};
\node at (-5,-1) (l5a)          {};
\node at (-2,-4) (l6)           {};
\node at (-1,-3) (cl1)          {};
\node at (0,-4)  (cl)           {};
\node at (1,-3)  (cr1)          {};
\node at (2,-4)  (r6)          {};
\node at (6,0)   (r5)          {};
\node at (7,-1)  (r4)          {};
\node at (8,0)   (r3)          {};
\node at (9,-1)  (r2)          {};
\node at (10,0)  (r1)          {};
\draw[to] (u) to (l1);
\draw[to] (l1) to (l2);
\draw[to] (l3) to (l2);
\draw[to] (l3) to (l4);
\draw[to] (l5) to (l4);
\draw[to] (l5) to (l5a);
\draw[to] (l5a) to (l6);
\draw[to] (cl1) to (l6);
\draw[to] (cl1) to (cl);
\draw[to] (cr1) to (cl);
\draw[to] (cr1) to (r6);
\draw[to] (r5) to (r6);
\draw[to] (r5)  to (r4);
\draw[to] (r3)  to (r4);
\draw[to] (r3)  to (r2);
\draw[to] (r1)  to (r2);
\draw[to] (u)  to (r1);
\draw[to,dotted] (u)  to node[anchor=north,sloped,pos=0.8] {$*$} (l3);
\draw[to,dotted] (u)  to node[anchor=north,sloped,pos=0.8] {$*$} (l5);
\draw[to,dotted] (u)  to node[anchor=north,sloped,pos=0.8] {$*$} (l5a);
\draw[to,dotted] (u)  to node[anchor=south,sloped,pos=0.8] {$*$} (cl1);
\draw[to,dotted] (u)  to node[anchor=south,sloped,pos=0.8] {$*$} (cr1);
\draw[to,dotted] (u)  to node[anchor=north,sloped,pos=0.8] {$*$} (r5);
\draw[to,dotted] (u)  to node[anchor=north,sloped,pos=0.8] {$*$} (r3);
\end{tikzpicture}
\caption{Additional property for the conversion version of Theorem~\ref{THM:main}.}
\label{FIG:fan}
\end{figure}
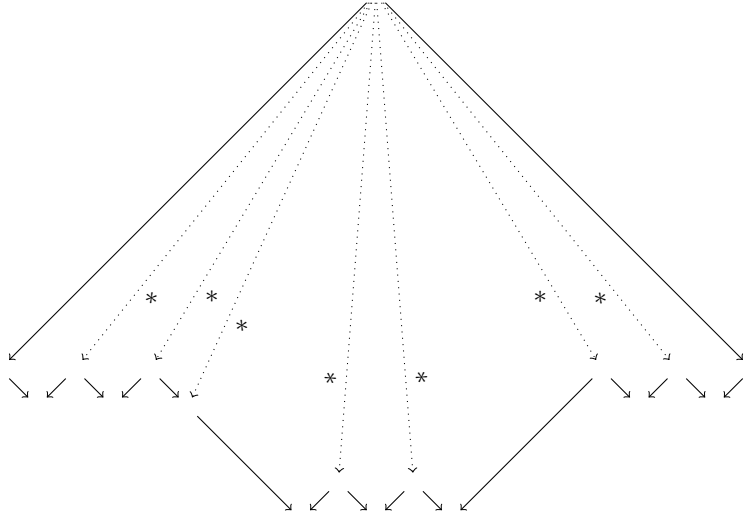

Regarding a variant of Theorem~\ref{THM:main} based on the conversion
version of decreasing diagrams the statement \quoted{As along a rewrite sequence
labels with respect to $\ell^\src$ never increase,~\ldots} in the above proof
is no longer appropriate, as rewrite sequences are replaced by conversions. 
Consequently, in contrast to the valley version, the source labeling might 
destroy decreasingness, as shown by the next example.

\begin{exa}[Example~\ref{ex:rl} continued]
\label{ex:rl:con}
Although the critical diagram in Figure~\ref{fig:ex:rl}\subref{fig:ex:rl:1} is 
decreasing for $\ell^i$ (cf.\ Example~\ref{ex:rl}) it is not decreasing for
$\ell^\src \times \ell^i$ as the label of the step
$\m{a}\xr[(\m{a},1)]{} \m{b}$ is not larger than
the label of the step $\m{b} \xl[(\m{c},0)]{} \m{c}$.
The problem is that $\m{a} \to^* \m{c}$ does not hold.
\end{exa}

To forbid the situation highlighted in Example~\ref{ex:rl:con},
the property that the labels of the conversions are smaller or equal to the source
of the local peak must be ensured.
\begin{defi}
\label{DEF:fan}
A diagram of the shape $t_1 \from s \to t_n$, $t_1 \conv t_2 \conv \cdots \conv t_n$
has the \emph{fan property} if $s \to^* t_i$ for $1\leqslant i\leqslant n$.
\end{defi}
The fan property is sketched in Figure~\ref{FIG:fan}, where the solid
arcs indicate the diagram and the dotted arcs the additional conditions.
Our formalization covers the following result, which is new in theory and \ISAFOR:
\begin{thm}[Conversion version of decreasing diagrams]
\label{THM:main:conv}
A left-linear TRS is confluent
if its duplicating rules terminate relative to its other rules and
all its critical peaks 
have local diagrams that both have the fan property and
are decreasing for the rule labeling.
\end{thm}
\proof
That $\ell^\src \times \ell^i$ is a labeling, is obtained as in the proof of
Theorem~\ref{THM:main}. The fan property ensures that a critical peak that
is decreasing for the rule labeling $\ell^i$ is also decreasing for
$\ell^\src \times \ell^i$. We conclude by an application of Corollary~\ref{COR:main}.
\qed

The following example demonstrates that the fan property is necessary
for Theorem~\ref{THM:main:conv} to be correct.
Note that the TRS in Example~\ref{ex:rl:con} is confluent and can
hence only motivate the fan property but cannot show its indispensability.
\begin{exa}
\label{exa:counter}
Consider the TRS $\RR$ consisting of the following rules,
where subscripts indicate labels for the rule labeling:
\begin{xalignat*}{5}
\f{a} &\to_2 \f{b} &
\f{f}(\f{a},\f{b}) &\to_1 \f{f}(\f{a},\f{a}) &
\f{f}(\f{b},\f{a}) &\to_1 \f{f}(\f{a},\f{a}) &
\f{f}(\f{a},\f{a}) &\to_1 \f{c} &
\f{g}(x) &\to_0 \f{f}(x,x)
\end{xalignat*}
Then $\RRdnd$ is easily seen to be terminating,
for example using 
the lexicographic path order with a quasi-precedence that equates
$\f{a}$ and $\f{b}$ and has $\f{g} > \f{f} > \f{c}$.
There are four critical peaks,
all of which are decreasing with respect to the given rule labeling:
\begin{xalignat*}{2}
\f{f}(\f{a},\f{b}) &\textstyle\xl[2]{} \f{f}(\f{a},\f{a}) \xr[1]{} \f{c} & &
  \textstyle \f{f}(\f{a},\f{b}) \xr[1]{} \f{f}(\f{a},\f{a}) \xr[1]{} \f{c} \\
\f{f}(\f{b},\f{a}) &\textstyle\xl[2]{} \f{f}(\f{a},\f{a}) \xr[1]{} \f{c} & &
  \textstyle \f{f}(\f{b},\f{a}) \xr[1]{} \f{f}(\f{a},\f{a}) \xr[1]{} \f{c} \\
\f{f}(\f{b},\f{b}) &\textstyle\xl[2]{} \f{f}(\f{a},\f{b}) \xr[1]{} \f{f}(\f{a},\f{a}) & &
  \textstyle \f{f}(\f{b},\f{b}) \xl[0]{} \f{g}(\f{b}) \xl[2]{} \f{g}(\f{a}) \xr[0]{} \f{f}(\f{a},\f{a}) \\
\f{f}(\f{b},\f{b}) &\textstyle\xl[2]{} \f{f}(\f{b},\f{a}) \xr[1]{} \f{f}(\f{a},\f{a}) & &
  \textstyle \f{f}(\f{b},\f{b}) \xl[0]{} \f{g}(\f{b}) \xl[2]{} \f{g}(\f{a}) \xr[0]{} \f{f}(\f{a},\f{a})
\end{xalignat*}
Nevertheless, $\RR$ is not confluent:
$\f{f}(\f{b},\f{b}) \from \f{f}(\f{a},\f{b}) \from \f{f}(\f{a},\f{a}) \to \f{c}$
is a conversion between two distinct normal forms.
Note that the local diagrams for the final two critical peaks violate
the fan property: $\f{g}(\f{a})$ is not reachable from $\f{f}(\f{a},\f{b})$
nor  $\f{f}(\f{b},\f{a})$.
\end{exa}

Finally, we remark that in the formalization Theorem~\ref{THM:main} is
obtained by instantiating
Theorem~\ref{THM:main:conv}. To this end, we formalized
that the fan property holds vacuously whenever the local diagram is a valley.
The direct proof of Theorem~\ref{THM:main} on page~\pageref{THM:main:proof} does
not have a correspondence in the formalization but it already conveys the proof idea
of the more complex proof of Theorem~\ref{THM:main:conv}.

\subsection{Certificates}
\label{APP:cert}
\begin{figure}[t]
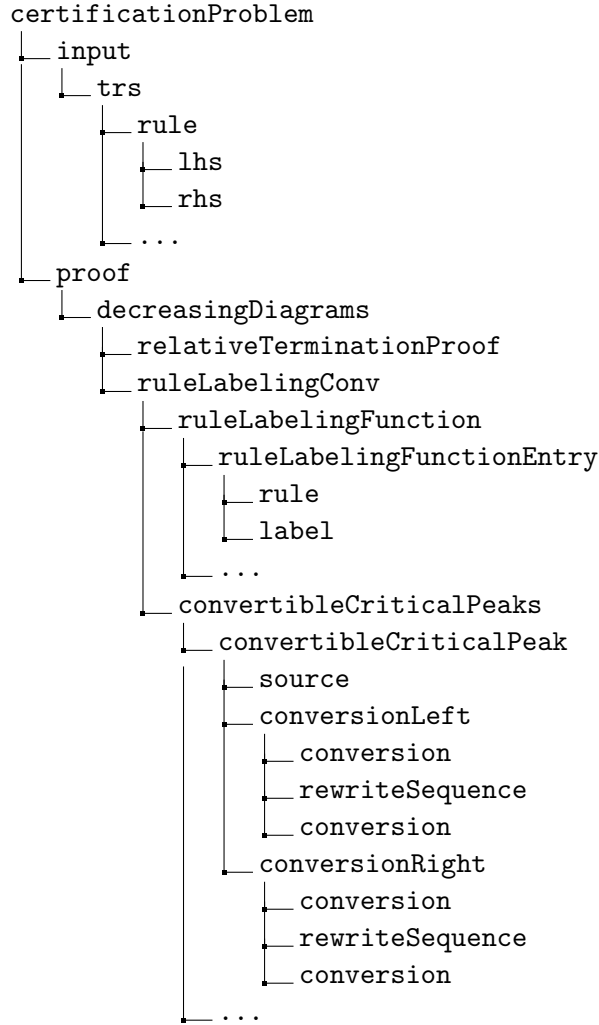

  \centering
  \begin{minipage}[t]{.6\linewidth}
    \dirtree{%
      .1 certificationProblem.
      .2 input.
      .3 trs.
      .4 rule.
      .5 lhs.
      .5 rhs.
      .4 ....
      .2 proof.
      .3 decreasingDiagrams.
      .4 relativeTerminationProof.
      .4 ruleLabelingConv.
      .5 ruleLabelingFunction.
      .6 ruleLabelingFunctionEntry.
      .7 rule.
      .7 label.
      .6 ....
      .5 convertibleCriticalPeaks.
      .6 convertibleCriticalPeak.
      .7 source.
      .7 conversionLeft.
      .8 conversion.
      .8 rewriteSequence.
      .8 conversion.
      .7 conversionRight.
      .8 conversion.
      .8 rewriteSequence.
      .8 conversion.
      .6 ....
    }
  \end{minipage}
\caption{Structure of a Rule Labeling Certificate in \CPF.}
\label{fig:cpfex}
\end{figure}

Next we discuss the design of the certificates for confluence proofs via 
Theorem~\ref{THM:main:conv}, i.e., how they are represented in \CPF,
and the executable checker to verify them.
The main structure of a certificate in \CPF consists of an input,
in our case the TRS whose confluence should be certified, and a proof.
For the proof, a minimal certificate
could just claim that the considered rewrite system can be shown decreasing via
the rule labeling. However, this is undecidable, even for locally confluent
systems~\cite{HM11}. Hence the certificate contains the following entries:
the TRS $\RR$, the index function~$i$,
(candidates for) the joining conversions for each critical
peak, an upper bound on the number of rewrite steps required to check the fan
property, and, in case $\RR$ is not right-linear, a relative termination proof
for $\RRdnd$.
The labels in the joining conversions are not required in the certificate, since
\CETA has to check, i.e., compute them anyway. The same holds for the critical
peaks.\footnote{Note that \ISAFOR already comes with a formalized
unification algorithm that was easy to reuse.}
Note that the (complex) reasoning required for parallel and variable peaks
does not pollute the certificates.
The outline of a certificate for a confluence proof
according to Theorem~\ref{THM:main:conv} is shown in
Figure~\ref{fig:cpfex}.\footnote{In Figure~\ref{fig:cpfex} some
boilerplate nodes and details are omitted---a full certificate in \CPF for Example~\ref{ex:rl}
is available at
\url{http://cl-informatik.uibk.ac.at/experiments/2016/rule_labeling/rule_labeling_conv.proof.xml}.}
Besides elements for structuring the certificate
(\texttt{decreasingDiagrams}, \texttt{ruleLabelingConv}), additions to \CPF,
for representing proofs via Theorem~\ref{THM:main:conv},
are the elements to represent the rule labeling
(\texttt{ruleLabelingFunction}) and the joining conversions
\texttt{convertibleCriticalPeaks}. Conversions and rewrite sequences themselves
were already representable in \CPF (\texttt{conversion}, \texttt{rewriteSequence})
and easy to reuse.
To check such a certificate \CETA performs the following steps:
\begin{enumerate}
\item Parse the certificate.
To parse the \CPF elements that were newly introduced
we extended \CETA's parser accordingly.
\item Check the proof for relative termination. Luckily, \CETA already supports a wide
range of relative termination techniques, so that here we just needed to make
use of existing machinery.
\item Compute all critical peaks of the rewrite system specified in the certificate.
\item For each computed critical peak, find and check
a decreasing joining conversion given in the certificate.
To check decreasingness (cf.\ Listing~\ref{lst:eld}) we require the decomposition of
the joining conversions to be explicit in the certificate.
As the confluence tools that generate certificates might use different renamings
than \CETA when computing critical pairs, the conversions given in the
certificate are subject to a variable renaming. Thus, after computing the
critical peaks, \CETA has to consider the joining conversions
modulo renaming of variables.
Then checking that they form a local diagram and that the labels are extended locally
decreasing is straightforward.
\item Check the fan property. The certificate contains an upper bound on the number of
rewrite steps required to reach the terms in the conversions from the source of the peak.
This ensures termination of \CETA when checking the existence of suitable rewrite sequences.
\end{enumerate}

\noindent \CETA also supports checking decreasingness using the valley version of decreasing
diagrams, i.e., certifying applications of Theorem~\ref{THM:main}.
In that case splitting the joining sequences in the certificate is
not required: for every critical peak just two rewrite sequences need to be
provided. \CETA can automatically find a split if one exists:
given two natural numbers $\alpha$ and $\beta$ and a sequence $\sigma$ of natural
numbers, is there a split $\sigma = \sigma_1\sigma_2\sigma_3$ such that
$\sigma_1 \subseteq \Vee\alpha$, $\sigma_2 \subseteq \Veq \beta$ with length of
$\sigma_2$ at most one, and $\sigma_3 \subseteq \Vee\alpha\beta$?
The checker employs a simple, greedy approach. That is, we pick the maximal
prefix of $\sigma$ with labels smaller $\alpha$ as $\sigma_1$. If the next label
is less or equal to $\beta$ we take it as $\sigma_2$
and otherwise we take the empty sequence for $\sigma_2$. Finally, the remainder of the
sequence is $\sigma_3$. A straightforward case analysis shows that this approach
is complete, i.e., otherwise no such split exists.

\section{Experiments}
\label{EXP:main}
In this section we compare the techniques from this paper with the
other confluence criteria supported by \CETA via an experimental evaluation.
For experiments we considered the \NUM TRSs
in Cops%
\footnote{The Confluence Problems database is available at \url{http://cops.uibk.ac.at}.}
that stem from the literature---this is also the pool from which the benchmarks for
CoCo are drafted---and used the confluence tool~\CSI~\cite{ZFM11b}
to obtain certificates in \CPF for confluence proofs.
\ACP~\cite{AYT09} can also produce confluence certificates in \CPF,
but at the moment they are a subset of the ones reported by \CSI.
All generated certificates have been certified by \CETA.
The largest certificate (for Cops~\#60) has 
760~KB and lists 182 candidate joins for showing the 34 critical peaks
decreasing. The certificate is checked within 1.1~seconds. We remark that
no confluence tool besides \CSI has solved Cops~\#60 so far, stressing the
importance of a certified proof.

\begin{table}[tb]
\centering
\begin{tabular}{l@{\qquad}rrrrr}
\toprule
method & success & CoCo\,'13 & CoCo\,'14 & CoCo\,'15 & CoCo\,'16 \\
\midrule
Knuth-Bendix                       & 27(+16) & \SUCC & \SUCC & \SUCC & \SUCC\\
(weak) orthogonality               &  4(+37) & \SUCC & \SUCC & \SUCC & \SUCC\\
strong closedness                  & 28(+18) & \SUCC & \SUCC & \SUCC & \SUCC\\
Lemma~\ref{LEM:rl}(\ref{LEM:rl:5}) & 44(+\phantom{1}6)  & \FAIL & \SUCC & \SUCC & \SUCC\\
parallel closedness                & 16(+36) & \FAIL & \FAIL & \FAIL & \SUCC\\
Theorem~\ref{THM:main}             & 49(+\phantom{1}7)  & \FAIL & \FAIL & \SUCC & \SUCC\\
redundant rules                    &  ---    & \FAIL & \FAIL & \SUCC & \SUCC\\
\midrule
$\sum$                             &         &  46   &  59   &   76  &   77\\
\bottomrule \addlinespace[\belowrulesep]
\end{tabular}
\caption{Experimental results for \NUM TRSs from Cops.}
\label{TAB:exp}
\end{table}

Next we elaborate on the impact of the new contributions.
Experimental results for various confluence criteria supported by \CETA
are shown in Table~\ref{TAB:exp}.%
\footnote{Details are available from
\url{http://cl-informatik.uibk.ac.at/experiments/2016/rule_labeling}.}
We track the progress of certification for confluence along CoCo.
The \CETA version from CoCo 2013 incorporated
(weak) orthogonality~\cite{R73},
Knuth-Bendix' criterion~\cite{KB70},
and Huet's result on strongly closed critical pairs~\cite{H80}.
As already employed for CoCo 2014,
we included the data for Theorem~\ref{THM:main} restricted to linear TRSs,
i.e., Lemma~\ref{LEM:rl}(\ref{LEM:rl:5}).
Due to the formalization described in this article, since CoCo 2015,
also Theorem~\ref{THM:main} is supported.
Since 2015 \CETA also supports a pre-processing technique for showing
confluence, namely addition and removal of redundant rules~\cite{NFM15}. The
idea of that technique is to add or remove rules that can be simulated by the
other rules of a TRS, which reflects confluence and often makes other criteria
applicable. The gain for the direct methods when combined with this technique is
shown in parentheses in the second column of Table~\ref{TAB:exp}.
Recently also Huet's result on parallel closed critical pairs~\cite{H80},
including Toyama's extension~\cite{T88}, was formalized and added to \CETA by
the first author~\cite{NM16}.
Note that~\CETA can also certify various methods for
non-confluence~\cite{NT14}, resulting in 36 certified non-confluence proofs.

\begin{figure}[t]
\centering
\begin{tikzpicture}
  \draw[rotate around={45:(0,1)}] (0,1) ellipse (1.3 and 2.5);
  \draw[rotate around={-45:(0,1)}] (0,1) ellipse (1.3 and 2.5);
  \draw[rotate around={45:(0,-1)}] (0,0.5) ellipse (1.5 and 2.5);
  \draw[rotate around={-45:(0,-1)}] (0,0.5) ellipse (1.5 and 2.5);
  \node at (-2.9,2.3) { KB };
  \node at (-1.5,3.3) { SC };
  \node at (1.5,3.3) { PC };
  \node at (2.9,2.3) { RL };
  \node at (-2.4,1.1) { 8 };
  \node at (-1.1,2.5) { 2 };
  \node at (1.1,2.5) { 5 };
  \node at (2.4,1.1) { 15 };
  \node at (0,-1.5) { 7 };
  \node at (1.5,-0.2) { 10 };
  \node at (0.6,-0.6) { 6 };
  \node at (-0.6,-0.6) { 1 };
  \node at (0.8,1) { 5 };
  \node at (0,0.3) { 5 };
\end{tikzpicture}
\caption{Overlap between solved Cops.}
\label{FIG:venn}
\end{figure}
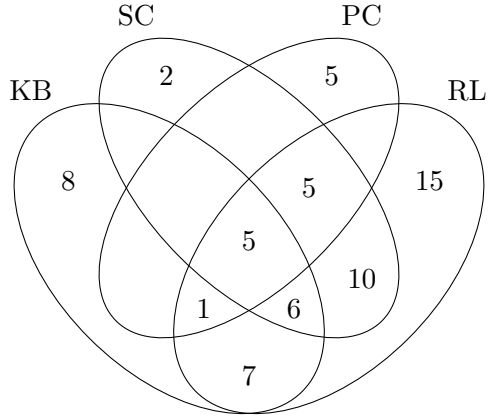

On our test-bed Theorem~\ref{THM:main}
can establish confluence of as many systems as all other direct methods (weak
orthogonality, Knuth-Bendix' criterion, strong and parallel closedness)
combined (49 systems), admits about 25\% increase in power (64 vs.\ 49) when
used in combination with them and by itself shows confluence of
about two thirds of all
systems that are certifiably confluent by {\CSI}+\CETA (77 vs.\ 49).
Figure~\ref{FIG:venn} shows the systems
solved by Knuth-Bendix' criterion (KB), strong closedness (SC), parallel closedness (PC),
and Theorem~\ref{THM:main} (RL) in relation to each other. (We omit
weak orthogonality and Lemma~\ref{LEM:rl}(\ref{LEM:rl:5}), since they
are subsumed by parallel closedness and Theorem~\ref{THM:main} respectively.)
Overall the decreasing diagrams based rule labeling technique turned out much stronger
than the other, simple, syntactic criteria. However, this increase in power comes at
a cost: the certificates are usually larger and much harder to check for a human,
making automatic, reliable certification a necessity.

We are not aware of a confluence tool that can yet produce certificates according
to Theorem~\ref{THM:main:conv}.
However, in the case of linear TRSs, \SAIGAWA supports the rule labeling
for the conversion version of decreasing diagrams~\cite{HM11}, which is covered
by Theorem~\ref{THM:main:conv} but not by Theorem~\ref{THM:main}. Producing
checkable certificates for these proofs would be an easy exercise for \SAIGAWA's
authors.

\section{Conclusion}
\label{CON:main}

Finally we discuss related work, comment on the existing 
formalization of rewriting in \ISAFOR, and conclude with a short summary.

\subsection{Related Work}
\label{CON:rel}

Formalizing confluence criteria has a long history in 
the $\lambda$-calculus.
Huet~\cite{H94} proved a stronger variant of the parallel moves lemma in
\COQ.
\ISABELLE/\HOL was used in~\cite{N01} to prove the Church-Rosser property
of $\beta$, $\eta$, and $\beta\eta$. For $\beta$-reduction the
standard Tait/Martin-L{\"o}f proof as well as
Takahashi's proof~\cite{T95} were formalized. The first mechanically
verified proof of the Church-Rosser property of $\beta$-reduction was
done using the Boyer-Moore theorem prover~\cite{S88}. The formalization
in Twelf~\cite{P92} was used to formalize the confluence proof of a
specific higher-order rewrite system in~\cite{S06}.

Next we discuss related work for term rewriting.
Newman's lemma (for abstract rewrite systems) and Knuth and Bendix' 
critical pair theorem (for first-order rewrite systems) have been
proved in~\cite{RRAHMM02} using \ACL.
An alternative proof of the latter in \PVS, following the higher-order
structure of Huet's proof, is presented in~\cite{GAR10}.
\PVS is also used in the formalization of the lemmas of Newman and
Yokouchi in~\cite{GAR08}. Knuth and Bendix' criterion has also been
formalized in \COQ~\cite{CCFPU11} and \ISABELLE/\HOL~\cite{ST13}.
The strong and parallel closedness conditions of
Huet~\cite{H80} have been formalized
by the first author in {\ISABELLE}/\HOL~\cite{NM16},
where reasoning similar to the one in Section~\ref{TRS:lp} is
used to close variable and parallel peaks. However, for
Huet's criteria it suffices to construct a common reduct while for our setting
every rewrite step has to be made explicit in order to compute the labels
and show local decreasingness.

\subsection{Assessment}

First we explain why Theorem~\ref{THM:main}
is an adequate candidate for a formalization.
On the one hand, regarding the aspect of automation, it is 
easily implementable as the
relative termination requirement can be outsourced to external (relative)
termination provers and the rule labeling heuristic has already been implemented
successfully~\cite{A10,HM10}. Furthermore, it is a powerful criterion as
demonstrated by the experimental evaluation in Section~\ref{EXP:main}.
On the other hand, regarding the aspect of formalization, it is 
challenging because it involves the combination of different labeling functions
(in the sense of~\cite{ZFM15}). Hence, in our formalization Theorem~\ref{THM:main}
is not established directly, but obtained as a corollary of more general results.
In particular Lemma~\ref{LEM:lex} is based on a more general result,
which allows different labeling functions to be combined lexicographically.
This paves the way for reusing the formalization described here 
when tackling the remaining criteria in~\cite{ZFM15},
which are based on more flexible combinations of labeling functions,
and use labelings besides the source labeling (Lemma~\ref{LEM:sn})
and the rule labeling (Lemma~\ref{LEM:rl}). For example, labels can also
be defined based on the path to a rewrite step, or the redex that is
being contracted. In order to certify the corresponding proofs, we will
also have to extend the \CPF format with encodings of those labelings.

The required
characterization of (closing) local peaks (cf.\ Figure~\ref{FIG:peaks}) provides
full information about the rewrite steps involved in the joining sequences. As
this characterization is the basis for many confluence criteria---not necessarily
relying on decreasing diagrams---this result aids future certification efforts.
We anticipate that the key result for closing variable
peaks for the left-linear case (cf.\ Section~\ref{TRS:lp}) does not rely
on the annotated version of parallel rewriting, but as~\cite{ZFM15} also
supports labelings based on parallel rewriting, the developed machinery
should be useful for targeting further confluence results from~\cite{ZFM15}.
Needless to say, parallel rewriting is handy on its own.
The formalization described in this article covers 
a significant amount of the results presented in~\cite{ZFM15}.
As explained, additional concepts 
(e.g., the annotated version of parallel rewriting)
were formalized with preparation of the remaining criteria in mind.
However, for some results that are not covered yet (e.g., persistency), we anticipate
that even formalizing the preliminaries requires significant effort.

Next we discuss the usefulness of existing formalizations for this work.
The existing machinery of \ISAFOR provided invaluable support.
We regard our efforts to establish an annotated version of
parallel rewriting not as a shortcoming of \ISAFOR, but as a useful extension to it.
On the contrary, we could employ many results from \ISAFOR without further
ado, e.g., completeness of the unification algorithm (to compute critical
peaks), plain rewriting (to connect parallel steps with single steps), and
the support for relative termination.
That Lemma~\ref{LEM:inst} occurred several times in \ISAFOR can be
traced to textbook proofs, e.g.~\cite{BN98}, where this result
is not made explicit either. Instead it is established in the scope of a larger proof
of the critical pair lemma.
Still, in later proofs the result is used
as if it would have been established explicitly. In \ISAFOR these proofs have
been duplicated, but as formalization papers typically come with code
refactoring these deficiencies have been fixed. Note that the duplicated 
proofs have actually never been published.
Ultimately our aim in the formalization was to follow paper proofs
as closely as possible. The benefit of this choice is that
this way, shortcomings
in existing proofs can be
identified and eradicated. As our formalization covers
and combines results from various sources, the notions used in the papers had
to be connected. As already mentioned, while different notations are typically
identified in paper proofs, in the formalization this step has to be made explicit. 
To avoid this drawback in the future our recommendation is to strive for more
standard notation, also in paper proofs.

Finally, differences to~\cite{ZFM15} are addressed.
The concepts of an L-labeling and an LL-labeling from~\cite{ZFM15}
have been generalized to the notion of a labeling \emph{compatible} with a TRS
while weak-LL-labelings are represented via \emph{weakly compatible} labelings
here.%
\footnote{The definitions of L-labelings and LL-labelings spell out the shape of the
standard joining valley for a variable peak for linear and left-linear TRSs,
respectively, and then impose restrictions on the occurring labels that
ensure compatibility.}
This admits the formulation of the abstract conditions such that a labeling
ensures confluence (cf.\ Corollary~\ref{COR:main}) independent from the TRS being
(left-)linear.
Furthermore we present a generalization of Theorem~\ref{THM:main} to
the conversion version of decreasing diagrams, namely Theorem~\ref{THM:main:conv}.

\subsection{Summary and Conclusion}

In this article we presented the formalization of a result establishing
confluence of left-linear term rewrite systems based on relative termination
and the rule labeling. While our formalization admits stronger results (in
order to prepare for further results from~\cite{ZFM15}), we 
targeted Theorem~\ref{THM:main}, whose statement (in contrast to its proof) does not
require the complex interplay of relative termination and the rule labeling.
Hence this criterion is easily implementable for automated confluence tools,
admitting the use of external termination provers.
Our formalization subsumes the (original) criterion for the rule labeling
(cf.\ Lemma~\ref{LEM:rl}\eqref{LEM:rl:5}), which is applicable to linear systems only.
Dealing with non-right-linear systems required an analysis of
non-right-linear variable peaks, and of the interplay with the
relative termination condition. Furthermore, whereas plain rule labeling
can be proved correct by decreasing diagrams, the involvement of the
source labeling means that extended decreasing diagrams are required.
Hence the proof of Theorem~\ref{THM:main} is significantly 
more involved than the one of Lemma~\ref{LEM:rl}\eqref{LEM:rl:5}.

Despite the fact that any confluence proof by the conversion version of decreasing diagrams
can be completed into a confluence proof by the valley version using the same
labels (cf.\ the proof of~\cite[Theorem~3]{vO08a}), conversions can be
(significantly) shorter than valleys~\cite[Example~8]{vO08a}. 
Regarding the conversion version of decreasing diagrams, in automated tools the
main obstacle is finding suitable conversions.  Albeit simple heuristics
(cf.~\cite[Section~4]{HM11})
have been proposed to limit the explosion in the search space when considering
conversions, most automated confluence provers still favor the valley
version. While those heuristics suffice for the rule labeling,
Theorem~\ref{THM:main:conv} shows that in a more complex setting,
conversions must satisfy additional restrictions, which make the
search for suitable conversions even more challenging. Maybe the recent
approach~\cite{Z16} to construct conversions can solve some of these challenges.

\section*{Acknowledgments}
We thank Christian Sternagel and Ren{\'e} Thiemann
for insightful discussion and the reviewers for helpful comments.

\bibliographystyle{plain}
\bibliography{references}

\end{document}